\documentclass[12pt,a4paper]{article}
\usepackage{amsmath,amssymb,url,cite,ifpdf,slashed,multirow,tabularx}

  \usepackage{graphicx}                    

\def\thefootnote{\ifnum\c@footnote>\z@\textasteriskcentered\@arabic\c@footnote\fi}
\makeatletter
\renewcommand{\footnoterule}{%
\kern-3\p@
\hrule width 0.4\columnwidth
\kern 2.6\p@}
\def\thefootnote{\ifnum\c@footnote>\z@\@arabic\c@footnote\fi}
\makeatother

\allowdisplaybreaks

\newcommand{\TeV}{\,{\rm TeV}}
\newcommand{\GeV}{\,{\rm GeV}}
\newcommand{\MeV}{\,{\rm MeV}}

\newcommand{\invfb}{\,{\rm fb^{-1}}}

\def\be{\begin{equation}}
\def\ee{\end{equation}}

\def\({\left(}
\def\){\right)}
\def\<{\langle}
\def\>{\rangle}

\newcommand{\Order}{\mathop{\mathcal{O}}}

\makeatletter
\newcommand{\@authornote}[2]{{\def\thefootnote{\fnsymbol{footnote}}\setcounter{footnote}{#1}#2\setcounter{footnote}{0}}}
\newcommand{\authornotemark}[1]{\@authornote#1{\addtocounter{footnote}{-1}\footnotemark}}
\newcommand{\authornotetext}[2]{\@authornote#1{\footnotetext{#2}}}
\makeatother

\begin{document}

\begin{titlepage}

\begin{flushright}
UT--13--34
\end{flushright}

\vskip 2.7cm
\begin{center}

{\Large \bf
Probing Bino Contribution 
to Muon  {\mbox{\boldmath $g-2$}}
}
\vskip .85in

{\large
\textbf{Motoi Endo}, 
\textbf{Koichi Hamaguchi},\\
\textbf{Teppei Kitahara}, 
\textbf{and}
\textbf{Takahiro Yoshinaga}\authornotemark{1}
}
\vskip 0.25in

\authornotetext{1}{Research Fellow of the Japan Society for the Promotion of Science}

{\em
Department of Physics,  University of Tokyo, Tokyo 113-0033, Japan
}

\end{center}
\vskip .75in

\begin{abstract}
We study SUSY models in which Bino contributions solve the muon $g-2$ anomaly.
The contributions are enhanced by large left-right mixing of the smuons.
However, it is constrained by the vacuum stability condition of the slepton--Higgs potential.
Therefore, there are upper bounds on masses of sleptons and Bino.
When the slepton soft masses are universal, 
the upper bound on the smuon mass becomes $330~(460)\GeV$ 
in order to solve the $g-2$ anomaly at the $1\sigma~(2\sigma)$ level.
It is within the reach of LHC and ILC.
If the stau is heavier than the smuon, the bound can be as large as $1.4~(1.9)\TeV$.
Such non-universal slepton mass spectrum generically predicts too large LFV/CPV.
We show that the models are expected to be probed by LHC/ILC and LFV/CPV complementarily in future.
\end{abstract}

\end{titlepage}
\setcounter{page}{2}

\section{Introduction} \label{sec:introduction}

The anomalous magnetic moment of the muon, $a_\mu = (g-2)_\mu/2$, (muon $g-2$) has been measured very precisely~\cite{g-2_bnl2010}. 
It is compared with the Standard Model (SM) prediction, and the latest result is 
\begin{equation}
 \Delta a_\mu \equiv a_\mu({\rm exp}) - a_\mu({\rm SM}) = (26.1 \pm 8.0) \times 10^{-10}, \label{eq:g-2_deviation}
\end{equation}
where Ref.~\cite{g-2_hagiwara2011} is referred to for contributions of the hadronic vacuum polarization,
and the hadronic  light-by-light contribution is from Ref.~\cite{Prades:2009tw}.
Similarly, Ref.~\cite{g-2_davier2010} provides $\Delta a_\mu = (28.6 \pm 8.0) \times 10^{-10}$. Therefore, there is discrepancy at more than $3\sigma$ confidence levels. It is noticed that the difference is as large as SM electroweak contributions, $a_\mu({\rm EW}) = (15.36 \pm 0.1) \times 10 ^{-10}$~\cite{Czarnecki:2002nt,Gnendiger:2013pva}.
If this is a signature of physics beyond the SM, and if the new physics exists in the TeV scale, new physics contributions to the muon $g-2$ are necessarily enhanced by some mechanisms, because they are naively estimated as $a_\mu({\rm NP}) \sim (\alpha _{\rm NP}/4\pi) \times (m_\mu^2 / m_{\rm NP}^2)$, which is required to be comparable to $a_\mu({\rm EW})$. 

The supersymmetry (SUSY) is a good candidate for such new physics models.
The model can provide sizable contributions to the muon $g-2$~\cite{lopez:1993vi,chattopadhyay:1995ae,moroi:1995yh}, which are enhanced by $\tan\beta = \langle H_u \rangle/\langle H_d \rangle$. 
The muon $g-2$ anomaly is solved if the superparticles (muonic sleptons, neutralinos and/or charginos) have a mass around $\mathcal{O}(100)\GeV$ for $\tan \beta = \mathcal{O}(10)$.
They are light enough to be produced at the LHC experiments. 
Recently, LHC phenomenology of the superparticles that are relevant for the muon $g-2$ has been studied in Ref.~\cite{Endo:2013bba}. It has been shown that they can be discovered in the near future in most of the parameter regions, especially in regions where the SUSY contributions to the muon $g-2$ are dominated by chargino--sneutrino diagrams. However, the searches rely on the assumption that the Wino is light. This assumption is not necessary to explain the muon $g-2$ discrepancy, when the SUSY contribution is mainly from Bino--smuon diagrams. If the Wino is heavy, collider searches differ significantly from those in Ref.~\cite{Endo:2013bba}. In this paper, we study searches for the models in which only the Bino and  the left- and right-handed smuons are light, 
while the other superparticles including the Wino are decoupled from the LHC sensitivity.

The Bino--smuon contribution is enhanced when the left-right mixing of the smuon is large. If the left-right mixing were allowed to be arbitrarily large, the superparticles could be extremely heavy while keeping the contribution to the muon $g-2$, thereby escaping any collider searches. However, too large left-right mixing spoils stability of the electroweak vacuum. Hence, the superparticle masses are bounded from above. We will show that, when staus and smuons have comparable masses to each others, 
upper bounds on the slepton masses are within the reach of LHC/ILC. If the staus are much heavier than smuons, the mass bounds are relaxed, while lepton flavor violations (LFV) and CP violations (CPV) generically become too large. Therefore, we will show that almost all the parameter regions can be tested in future complementarily by LHC/ILC and LFV/CPV, if the muon $g-2$ anomaly is solved by the Bino and the left- and right-handed smuons. 

The rest of the paper is organized as follows.
The mass spectrum is provided in Sec.~\ref{sec:setup}. The SUSY contributions to the muon $g-2$ and the vacuum stability conditions are explained in Sec.~\ref{sec:pheno}. In Sec.~\ref{sec:search}, experimental searches are studied. Sec.~\ref{sec:conclusion} is devoted to the conclusion.


\section{Mass spectrum} \label{sec:setup}

We assume that the SUSY contributions to the muon $g-2$ are dominated by the Bino--smuon contribution. The Bino and the left- and right-handed smuons contribute to the diagram. It is enhanced by the left-right mixing of the smuon, which is determined by the muon Yukawa coupling constant, the Higgsino mass parameter, $\mu$, and $\tan\beta$. Details will be discussed in the next section. Since too large $\tan\beta$ spoils perturbativity of the down-type Yukawa interactions, $\mu$ is favored to be large to explain the muon $g-2$ discrepancy \eqref{eq:g-2_deviation}. Thus, we focus on large Higgsino mass regions. The Winos are also supposed to be decoupled.

The collider searches depend on the slepton mass spectrum. The left- and right-handed selectron masses are assumed to be degenerate with those of the smuons, respectively. The following conclusion does not depend on this assumption.\footnote{
If the selectron masses are non-universal in the first two generations, the LFV bound is severe. See the discussions in Sec.~\ref{sec:non-univ}.
} 
On the other hand, the vacuum stability condition and the LFV/CPV bounds depend on the stau masses (see Sec.~\ref{sec:vac} and \ref{sec:non-univ}). 

All colored superparticles are set to be very heavy. In fact, none of them have been discovered at LHC. The Higgs boson mass of $126\GeV$ favors the scalar top masses to be $\Order(10-100)\TeV$, if the trilinear coupling of the top squark is not large. Similarly, the heavy Higgs bosons of the two Higgs doublets are assumed to be heavy. In this paper, all of them are considered to be decoupled. 

Consequently, we consider the low-energy effective theory, in which only the following superparticles are light,
\begin{equation}
  \tilde B,~\tilde{\ell}_L,~\tilde{\ell}_R.
  \label{eq:particles}
\end{equation}
Here, $\tilde{\ell}$ denotes the selectron and the smuon (and the stau, depending on the mass spectrum). In addition, the Higgsinos can contribute to the effective Lagrangian, if they are light due to the vacuum stability condition. The model parameters are 
\begin{equation}
M_1, m_{\tilde\ell_L}^2, m_{\tilde\ell_R}^2, m_{\tilde\ell_{LR}}^2.
\label{eq:parameters}
\end{equation}
Here, $M_1$ is the Bino mass. Since the Higgsinos (and the Wino) are heavy, the lightest neutralino is almost composed of the Bino. On the other hand, $m_{\tilde\ell_L}^2$ and $m_{\tilde\ell_R}^2$ are soft SUSY-breaking masses of the left- and right-handed sleptons, respectively. 
$m_{\tilde\ell_{LR}}^2$ is off-diagonal components of the slepton mass matrices. 
This controls the observables and constraints that are studied in this paper. 

Before proceeding to the phenomenology, let us comment on the left-right mixing of the smuon. It includes the scalar trilinear coupling of the muon, $A_\mu$, as well as $\mu$ and $\tan\beta$. The Bino--smuon contribution could be enhanced by $A_\mu$ with $\mu\tan\beta$ kept small. However, this requires $A_\mu$ to be extraordinary large. If the trilinear coupling is universal among the matter scalar fermions, this implies extremely large trilinear coupling for the stop sector, resulting in either too large Higgs boson mass or rapid decays of our electroweak vacuum into charge/color breaking vacua. In this paper, $A_\mu$ is set to be zero for simplicity, and the left-right mixing of the smuon is determined by $\mu$ and $\tan\beta$.


\section{Muon {\mbox{\boldmath $g-2$}}  and Vacuum Stability } \label{sec:pheno}

\subsection{Muon {\mbox{\boldmath $g-2$}} } \label{sec:g-2}

\begin{figure}[tbp]
 \begin{center}
 \includegraphics[width=8cm]{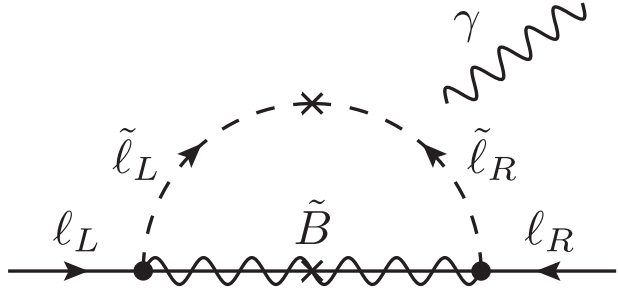}
 \end{center}
 \caption{The Bino--smuon contribution to the magnetic dipole operators.}
 \label{fig:bino}
\end{figure}

The Bino--smuon contribution to the muon $g-2$ is represented by Fig.~\ref{fig:bino}, where the lepton $\ell$ is the muon. In the mass insertion approximation, it is estimated as (cf.~\cite{moroi:1995yh})
\begin{align}
 a_{\mu}({\rm SUSY}) 
 &= 
 -(1 + \delta^{\rm 2loop}) \frac{\alpha_Y}{4\pi}
 \frac{ m_\mu M_1 m_{\tilde\mu_{LR}}^2 }{m^2 _{\tilde{\mu}_{L}}m^2 _{\tilde{\mu}_{R}}}\,
 f_N \left( \frac{m^2_{\tilde{\mu}_{L}}}{M^2_1},\frac{m^2 _{\tilde{\mu}_{R}}}{M^2_1} \right)
 \notag \\ 
 &= 
 \frac{1 + \delta^{\rm 2loop}}{1 + \Delta_\mu} \frac{\alpha_Y}{4\pi}
 \frac{ m_{\mu }^2  M_1 \mu }{m^2 _{\tilde{\mu}_{L}}m^2 _{\tilde{\mu}_{R}}} \tan \beta 
 \cdot f_N \left( \frac{m^2_{\tilde{\mu}_{L}}}{M^2_1},\frac{m^2 _{\tilde{\mu}_{R}}}{M^2_1} \right)
 \notag \\
 &\simeq
1.5\times 10^{-9} ~\frac{1 + \delta^{\rm 2loop}}{1 + \Delta_\mu}
\left(\frac{\tan\beta}{10}\right)
\left(\frac{(100\GeV)^2}{m_{\tilde{\mu }_L}^2 m_{\tilde{\mu }_R}^2/M_1\mu }\right)
\left(\frac{f_N}{1/6}\right),
\label{eq:gminus2neut}
\end{align}
at the leading order of $\tan \beta$. The loop function $f_N(x,y)$ is defined as
\begin{equation}
 f_N(x,y) = xy \left[ \frac{-3 + x + y + xy}{(x-1)^2(y-1)^2} + \frac{2x \log x}{(x-y)(x-1)^3} - \frac{2y \log y}{(x-y)(y-1)^3} \right], \label{eq:fN}
\end{equation}
which satisfies $f_N(1,1)=1/6$ and $0\le f_N(1,1)\le 1$.
It is noticed that the contribution is proportional to the left-right mixing, 
\begin{align}
m_{\tilde\ell_{LR}}^2 
= - Y_\ell\, v_u \mu
= - \frac{m_\ell}{1+\Delta_\ell} \mu \tan\beta,
\label{eq:LRmixing}
\end{align}
where $v_u$ is the vacuum expectation value (VEV) of the up-type Higgs field. Note that the heavy Higgs bosons are decoupled. In the last equation, $\Delta_\ell$ is a correction to the lepton Yukawa coupling constant~\cite{Marchetti:2008hw}. It appears when the lepton Yukawa coupling is matched to the physical lepton mass, $m_\ell$, or the Yukawa coupling in SM. In the low-energy effective theory, it becomes
\begin{align}
\Delta_\ell = \frac{\alpha_Y}{4\pi} M_1 \mu\tan\beta 
\cdot I(M_1^2, m_{\tilde\ell_L}^2, m_{\tilde\ell_R}^2),
\label{eq:Yukawa}
\end{align}
where Higgsino diagrams are discarded, because they are suppressed in large Higgsino mass regions. Also, the terms that are not enhanced by $\tan\beta$ are omitted. The loop function is defined as
\begin{align}
I(a,b,c) = -\frac{ab\ln a/b + bc\ln b/c + ca\ln c/a}{(a-b)(b-c)(c-a)},
\end{align}
which satisfies $f(a,a,a)=1/2a$. In particular, when $\mu \tan\beta$ is very large, $\Delta_\mu$ becomes as large as or larger than $\Order(0.1-1)$.

The correction $\delta^{\rm 2 loop}$ denotes leading contributions of two loop diagrams. It is estimated as
\begin{align}
1 + \delta^{\rm 2 loop} = 
\left(1 - \frac{4\alpha}{\pi} \ln \frac{m_{\rm soft}}{m_\mu} \right)
\left[1 + \frac{1}{4\pi}
  \left( 2\alpha_Y\Delta b + \frac{9}{4}\alpha_2 \right) \ln \frac{M_{\rm soft}}{m_{\rm soft}} 
\right].
\label{eq:2-loop}
\end{align}
In the right-hand side, the first bracket is QED corrections to the muon $g-2$~\cite{vonWeitershausen:2010zr}, or renormalization group contributions to the effective operator of the magnetic dipole operator from the (smuon or Bino) soft mass scale, $m_{\rm soft}$, to the muon mass scale. In the numerical analysis, $m_{\rm soft}$ is chosen to be the smuon mass. This correction is $\sim 10\%$. Non-logarithmic terms evaluated in Ref.~\cite{vonWeitershausen:2010zr} are found to be very small in the parameter regions of this paper. 

The second bracket in Eq.\eqref{eq:2-loop} is corrections to the Bino couplings with the smuons. When SUSY is exact, the gaugino coupling is equal to the gauge coupling constant. This equality is violated after heavy superparticles are decoupled at a scale, $M_{\rm soft}$. The Bino--muon--smuon interactions are
\begin{align}
 \mathcal{L}_{\text{int}} = -\frac{1}{\sqrt{2}}\tilde{g}_L\, \overline{\tilde{B}} \mu _L\, \tilde{\mu }^*_L + \sqrt{2} \tilde{g}_R\, \overline{\tilde{B}} \mu _R\, \tilde{\mu}_R^* + {\rm h.c.},
\end{align}
where the coefficients are 
\begin{align}
\tilde g_L &= g_Y + \delta\tilde g_L \simeq 
g_Y 
\left[1 + \frac{1}{4\pi}
  \left( \alpha_Y\Delta b + \frac{9}{4}\alpha_2 \right) \ln \frac{M_{\rm soft}}{m_{\rm soft}} 
\right], \label{eq:gL} \\
\tilde g_R &= g_Y + \delta\tilde g_R \simeq
g_Y 
\left[1 + \frac{\alpha_Y}{4\pi}
  \Delta b\, \ln \frac{M_{\rm soft}}{m_{\rm soft}} 
\right]. \label{eq:gR}
\end{align}
Here, the terms of $\alpha_Y$ are corrections to the Bino self-energy, which are SUSY analog of the oblique corrections~\cite{Nojiri:1996fp,Nojiri:1997ma,Cheng:1997sq}. The U(1)$_Y$ gauge coupling constant, $g_Y$, is evaluated at $m_{\rm soft}$, and the corrections are represented by the difference between the beta functions of the U(1)$_Y$ gauge and Bino couplings. In our setup, the SM particles and the sleptons contribute to the beta functions. Thus, the coefficient becomes $\Delta b = 41/6 - n_{\rm slepton}$, where $n_{\rm slepton}$ is number of the generations of light sleptons; for instance, $n_{\rm slepton} = 3$ if all the sleptons are light, and $n_{\rm slepton} = 2$ when the staus are decoupled. 
On the other hand, the term of $\alpha_2$ is non-oblique corrections after the Wino decoupled (cf. Ref.~\cite{Hikasa:1995bw}). The weak boson couples only to the left-handed (s)leptons at the Bino--muon--smuon vertices. Since the Bino--smuon contribution to the muon $g-2$ includes both $\tilde g_L$ and $\tilde g_R$, the loop correction \eqref{eq:2-loop} is obtained. It yields a correction of $5-10\%$ for $M_{\rm soft} = 10-100\TeV$ with $m_{\rm soft} \sim 100\GeV$. In the analysis, only the logarithmic contributions are included. Non-logarithmic terms are expected to be suppressed. 

\begin{figure}[tbp]
 \begin{center}
 \includegraphics[width=7cm]{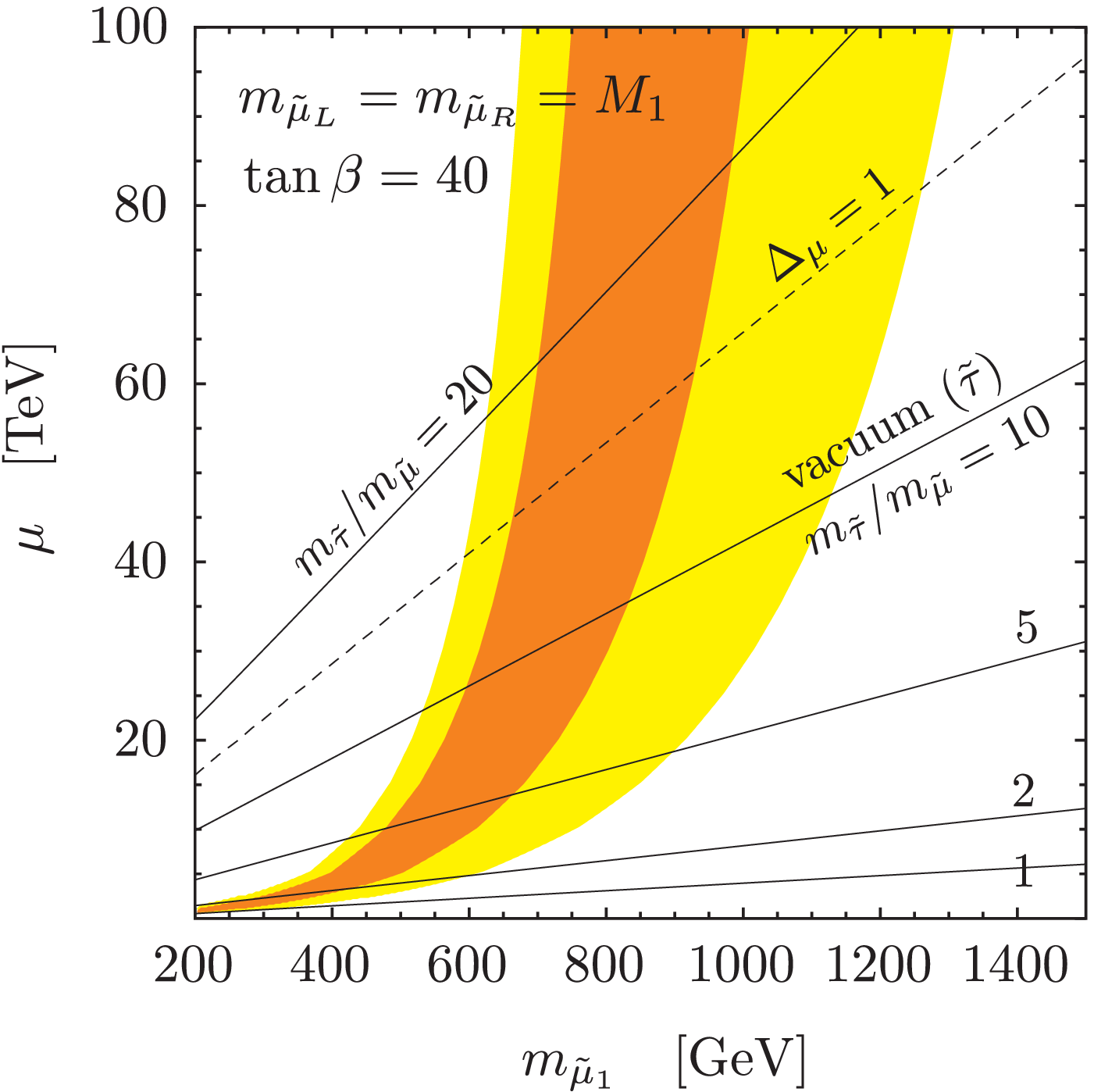}\hspace*{5mm}
 \includegraphics[width=7cm]{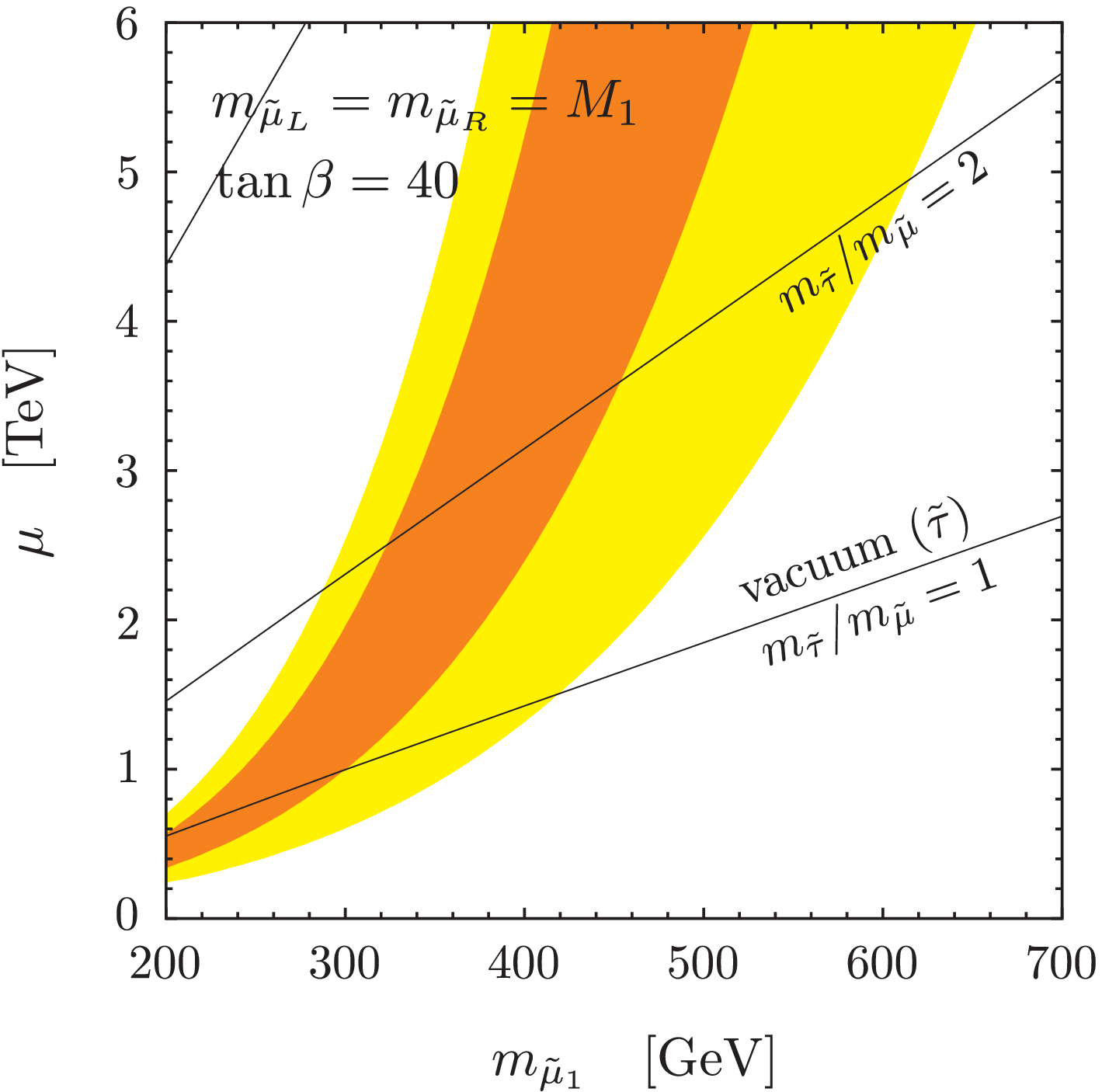}
 \end{center}
 \caption{
The SUSY contributions to the muon $g-2$ and the vacuum stability bounds are shown. 
In the orange (yellow) regions, the muon $g-2$ discrepancy \eqref{eq:g-2_deviation} is explained at the $1 \sigma$ $(2 \sigma)$ level. 
The black solid lines represent the upper bound on $\mu$ from the vacuum stability bound of the stau-Higgs potential. The stau masses are set to be $m_{\tilde{\tau}}/m_{\tilde{\mu}}=1, 2, 5, 10$ and 20 from bottom to top.
Above the black dashed line, $\Delta_{\mu}$ becomes larger than unity. 
The parameters are $M_1 = m_{\tilde\mu_L} = m_{\tilde\mu_R}$, $\tan\beta = 40$ and $M_{\rm soft} = 30\TeV$. 
In the right panel, a part of the parameter region of the left panel is magnified.
}
\label{fig:gminus2_stausmubound}
\end{figure}

In Eq.~\eqref{eq:2-loop}, the corrections that can be as large as or larger than $\Order(10)\%$ are included. Other two loop contributions are unknown or expected to be small. In Refs.~\cite{Heinemeyer:2003dq,Feng:2008nm}, SUSY corrections to the SM one-loop diagrams are evaluated. When the superparticles except those of \eqref{eq:particles} are decoupled, the contributions in Refs.~\cite{Heinemeyer:2003dq,Feng:2008nm} are negligibly small. Other corrections that have not yet been estimated include electroweak and SUSY two-loop contributions to SUSY one-loop diagrams. They might provide $\delta a_\mu \sim 10^{-10}$, according to Ref.~\cite{Stockinger:2006zn}. In addition, non-logarithmic corrections to $\delta^{\rm 2 loop}$ could be a few percents of the SUSY one-loop contributions, similarly to the discussions in Ref.~\cite{Stockinger:2006zn}. 

Apart from the Bino--smuon diagrams, there are other one-loop contributions to the muon $g-2$. The Bino--Higgsino--smuon contribution can be $\lesssim \Order(10^{-10})$ when the Higgsinos are light due to the vacuum stability bound (Sec.~\ref{sec:vac}). It is included in the numerical analysis for completeness.\footnote{
This contribution can dominate the SUSY contributions to the muon $g-2$, when $\mu$ is small while decoupling the Wino. Since they are enhanced only by $\tan\beta$, superparticles are required to be light to explain \eqref{eq:g-2_deviation}. They are detectable in colliders. 
In particular, the Higgsino production can be significant. 
}
On the other hand, the chargino--muon sneutrino contributions are less than $\Order(10^{-11})$ for $M_2 > 10\TeV$, i.e., negligible. 

In Fig.~\ref{fig:gminus2_stausmubound}, contours of the SUSY contributions to the muon $g-2$ are shown. The horizontal and vertical axises are the lightest smuon mass, $m_{\tilde\mu_1}$, and $\mu$, respectively. The parameters are set as $M_1 = m_{\tilde\mu_L} = m_{\tilde\mu_R}$, $\tan\beta = 40$ and $M_{\rm soft} = 30\TeV$. In the orange (yellow) regions, the SUSY contributions explain the muon $g-2$ discrepancy \eqref{eq:g-2_deviation} at the $1 \sigma$ $(2 \sigma)$ level. It is found that they are enhanced by large $\mu$, and the smuon masses can be $1\TeV$ for $\mu=\Order(10-100)\TeV$. This is contrasted to the chargino--muon sneutrino contributions to the muon $g-2$, where $\mu$ is favored to be small~\cite{Endo:2013bba}. On the other hand, detailed dependences on the superparticle mass spectrum are determined by the loop function \eqref{eq:fN} and the vacuum stability condition. They will be discussed in the next subsection.

\subsection{Vacuum stability} \label{sec:vac}

As shown in Sec.~\ref{sec:g-2}, the Bino--smuon contribution to the muon $g-2$ is enhanced by a large left-right mixing of the smuon. However, too large mixing spoils the stability of the electroweak vacuum. The trilinear coupling of the sleptons and the SM-like Higgs boson is given by
\begin{align}
V & \simeq 
\frac{1}{\sqrt{2}v}m_{\tilde\ell_{LR}}^2 \tilde{\ell}_{L}^* \tilde{\ell}_{R} h^0 + {\rm h.c.} \notag \\
& =
-\frac{m_{\ell}}{\sqrt{2}v(1+\Delta_\ell)}\mu\tan\beta 
\cdot\tilde{\ell}_{L}^* \tilde{\ell}_{R} h^0 + {\rm h.c.},
\label{eq:vacuum_bound}
\end{align}
where $v \simeq 174\GeV$ is the Higgs VEV. As the trilinear coupling increases, disastrous charge-breaking minima in the scalar potential become deeper, and our electroweak vacuum could decay to them. By requiring that the lifetime of the electroweak vacuum should be longer than the age of the Universe, $m_{\tilde\ell_{LR}}^2$ is constrained. 

The vacuum stability conditions have been studied. The fitting formula of the stability condition is obtained as 
\begin{align}
\left|m_{\tilde\ell_{LR}}^2 \right| 
&\leq
\eta_\ell 
\bigg[
1. 01  \times 10^2 \GeV \sqrt{m_{\tilde{\ell}_L} m_{\tilde{\ell}_R}} 
+ 1.01 \times 10^2 \GeV  (m_{\tilde{\ell}_L} + 1.03 m_{\tilde{\ell}_R}) 
\notag \\
-2.27 \times 
&
10^4\GeV^2
+ \frac{2.97 \times 10^6\GeV^3}{m_{\tilde{\ell}_L} + m_{\tilde{\ell}_R}} 
- 1.14 \times 10^8\GeV^4 
  \left( \frac{1}{m^2_{\tilde{\ell}_L}} +  \frac{0.983}{m^2_{\tilde{\ell}_R}} \right) 
\bigg]. \label{eq:kyb}
\end{align}
This is consistent with the result of Ref.~\cite{Kitahara:2013lfa} in a case of the stau--Higgs potential.
The vacuum decay rate is evaluated by the bounce method~\cite{Coleman:1977py}. The public package \texttt{CosmoTransitions 1.0.2}~\cite{Wainwright:2011kj} is used for the numerical evaluation. The scalar potential of the left- and right-handed sleptons and the SM-like Higgs boson is analyzed at the zero temperature.\footnote{
Thermal corrections can change the transition rate. They could be significant when the staus are light, for instance, for $m_{\tilde\tau_1} \lesssim 200\GeV$~\cite{Endo:2010ya}. 
}
The Higgs potential is set to reproduce the mass of $126\GeV$. Only the renormalizable terms are taken into account. Higher-dimensional terms depend on the superparticles that decouple at $M_{\rm soft}$. Since $M_{\rm soft}$ is very large, their contributions to \eqref{eq:kyb} are considered to be small. The fitting formula reproduces results of \texttt{CosmoTransitions} at better than the 1\% level. 

The fitting formula (\ref{eq:kyb}) is universal for all the slepton flavors up to corrections, $\eta_\ell \sim 1$. Given the soft scalar masses, the bound is independent of the flavor except through Yukawa interactions in the quartic terms of the scalar potential. Since the quartic terms are dominated by gauge interactions, $\eta_\ell$ changes little around unity. It also depends on $\tan\beta$ when the Yukawa coupling is large. In practice, $\eta_\tau = 1$ is set for the stau with $\tan\beta/(1+\Delta_\tau) = 70$. Numerical result of $\eta_\tau$ for the stau with various $\tan\beta$ is found in Fig.~2 of Ref.~\cite{Kitahara:2013lfa}. For instance, it is $\eta_\tau \simeq 0.94$ for $\tan\beta/(1+\Delta_\tau) = 40$. On the other hand, we obtain $\eta_\ell \simeq 0.88$ for the smuons and the selectrons, which is independent of the flavor and $\tan\beta$ because of small Yukawa couplings.

The most severe constraint on $\mu\tan\beta$ is obtained from the stability condition of the stau-Higgs potential. This is because the left-right mixing is proportional to the Yukawa coupling. In Fig.~\ref{fig:gminus2_stausmubound}, the upper limits are shown for $\tan\beta = 40$. Combined with the muon $g-2$, the smuon masses are bounded from above for given stau masses. For $m_{\tilde e} = m_{\tilde\mu} = m_{\tilde\tau}$, the lightest smuon is limited to be $m_{\tilde{\mu}_1}\lesssim 300~(420)\GeV$ at the $1\sigma$ ($2\sigma$) level of the muon $g-2$. 

\begin{figure}[tbp]
 \begin{center}
 \includegraphics[width=7cm]{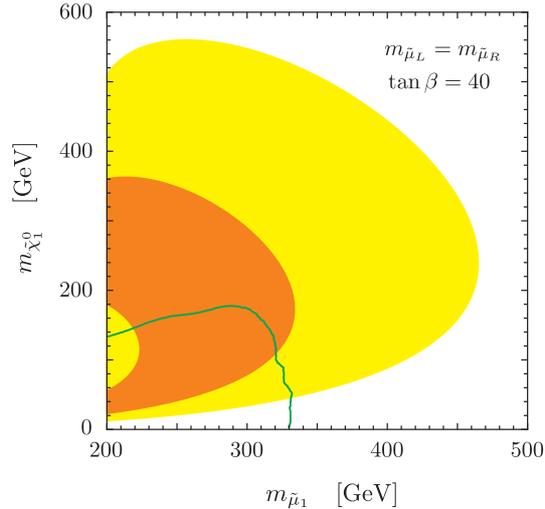}
 \caption{The SUSY contributions to the muon $g-2$ as a function of the lightest smuon mass, $m_{\tilde\mu_1}$, and the lightest neutralino mass, $m_{\tilde\chi^0_1}$.
 In the orange (yellow) regions, the muon $g-2$ discrepancy \eqref{eq:g-2_deviation} is explained at the $1 \sigma$ $(2 \sigma)$ level.  The left-right mixing is maximized under the vacuum stability condition. 
 The parameters are $m_{\tilde\ell_L} = m_{\tilde\ell_R}$, $\tan\beta = 40$ and $M_{\rm soft} = 10\TeV$. The stau soft masses are equal to those of the selectrons and smuons.
 The region below the green line is excluded by LHC.}
 \label{fig:MSL-M1}
 \end{center}
\end{figure}

The SUSY contributions to the muon $g-2$ depend on superparticle mass spectra mainly through the loop function \eqref{eq:fN}. In Fig.~\ref{fig:MSL-M1}, contours of the muon $g-2$ are displayed for various smuon and neutralino masses. In the orange (yellow) regions, the muon $g-2$ discrepancy is explained at the $1\sigma$ ($2\sigma$) level. The left-right mixing is maximized with satisfying the vacuum stability condition. Here, the stau soft mass is supposed to be degenerate with that of the smuon. It is found that the lightest smuon mass can be $330~(460)\GeV$ at the $1\sigma$ ($2\sigma$) level of the muon $g-2$, when the lightest neutralino mass, $m_{\tilde\chi^0_1}$, is almost a half of $m_{\tilde{\mu}_1}$. Note that $m_{\tilde\chi^0_1}$ is almost equal to $M_1$. On the other hand, $m_{\tilde\chi^0_1} \simeq 560\GeV$ is realized for $m_{\tilde{\mu}_1} \simeq 250\GeV$ at the $2\sigma$ level of the muon $g-2$. 

For a fixed value of the lightest smuon mass, the SUSY contributions to the muon $g-2$ is maximized when $m_{\tilde\mu_L}$ is equal to $m_{\tilde\mu_R}$. Eq.~\eqref{eq:gminus2neut} depends on the left- and right-handed smuon masses as $f_N(x,y)/xy$ with $x = m_{\tilde\mu_L}^2/M_1^2$ and $y = m_{\tilde\mu_R}^2/M_1^2$. When $xy$ is fixed, it is maximized for $x=y$ and rapidly decreases for $x \neq y$. On the other hand, the vacuum stability condition \eqref{eq:kyb} is relaxed for $m_{\tilde\ell_L} \neq m_{\tilde\ell_R}$. Since the former dependence is stronger than the latter, the SUSY contributions to the muon $g-2$ become largest when $m_{\tilde\mu_L}$ is equal to $m_{\tilde\mu_R}$ for given $M_1$ and $\tan\beta$. In other words, the orange/yellow regions in Fig.~\ref{fig:MSL-M1}, i.e., the parameter regions favored by the muon $g-2$, shrink toward smaller superparticle masses for $m_{\tilde\mu_L} \neq m_{\tilde\mu_R}$. 

\begin{figure}[tb]
 \begin{center}
 \includegraphics[width=7cm]{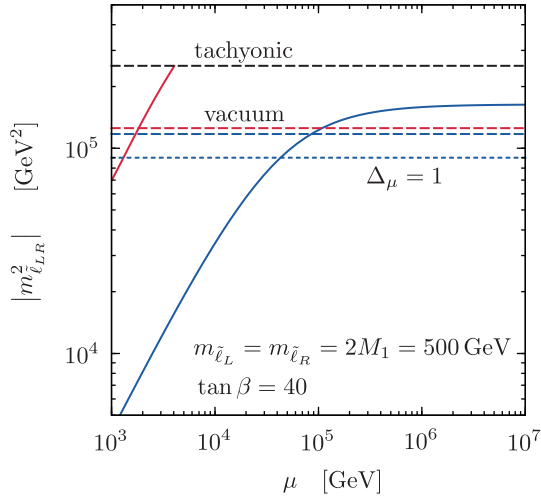}
 \caption{The left-right mixing of the smuon (blue solid) and the stau (red solid) as a function of $\mu$.
 The blue (red) dashed line represents the vacuum meta-stability bound with $\eta = 0.88 ~(0.94)$.
 Above the black dashed line, sleptons become tachyonic. 
 Below the blue dotted line, $\Delta_{\mu} < 1$ is satisfied.
 The parameters are $m_{\tilde\ell_L} = m_{\tilde\ell_R} = 2M_1 = 500\GeV$, $\tan\beta = 40$ and $M_{\rm soft} = 30\TeV$.}
 \label{fig:LRmixing}
 \end{center}
\end{figure}

The vacuum stability bound from the stau-Higgs potential is relaxed for heavier staus. When they are decoupled, the constraint almost disappears, and larger $\mu$ is allowed. As $\mu$ increases, the left-right mixing \eqref{eq:LRmixing} approaches to a maximal value. It is determined by the soft scalar masses as
\begin{align}
m_{\tilde\ell_{LR}}^2 = - \frac{m_\ell}{1+\Delta_\ell} \mu \tan\beta 
~\to~ 
- \frac{4\pi}{\alpha_Y} \frac{m_\ell}{M_1} \frac{1}{I(M_1^2, m_{\tilde\ell_L}^2, m_{\tilde\ell_R}^2)} ~~~( \text{for}~\mu \to \infty ).
\label{eq:mLRmax}
\end{align}
Note that the correction $\Delta_\ell$ is much larger than unity in this limit. In Fig.~\ref{fig:LRmixing}, the left-right mixings of the smuon and the stau, $|m_{\tilde\ell_{LR}}^2|$, are plotted as a function of $\mu$, by the blue and red solid lines, respectively. Since $m_{\tilde\ell_{LR}}^2$ is proportional to the Yukawa coupling, $|m_{\tilde\tau_{LR}}^2|$ is larger than $|m_{\tilde\mu_{LR}}^2|$. When the staus are light, $\mu$ is bounded from above by the stability condition of the stau--Higgs potential (red dashed line). For heavier staus, the red dashed line (as well as the stau tachyonic bound or the black dashed line) is lifted up, and larger $\mu$ is allowed. 

Even if the staus are decoupled, the left-right mixing is limited by the vacuum stability condition of the smuon-Higgs potential. 
When $M_1$ is smaller than the smuon mass with $m_{\tilde\mu_L} = m_{\tilde\mu_R}$, 
the vacuum constraint (\ref{eq:kyb}) is more severe than Eq. (\ref{eq:mLRmax}).
It can be estimated that the lightest smuon can be as large as $1.4~(1.9)\TeV$ at the $1\sigma~(2\sigma)$ level of the muon $g-2$.\footnote{
In this limit, the situation is essentially the same as the soft Yukawa coupling models~\cite{Borzumati:1999sp,Crivellin:2010ty}. In this paper, the vacuum stability condition is analyzed seriously and is found to restrict the SUSY contributions to the muon $g-2$, resulting in tighter bound on the smuon mass than Ref.~\cite{Crivellin:2010ty}.
}
Since it is very weak, the bound does not appear in the parameter range of Fig.~\ref{fig:gminus2_stausmubound}. 

It is noticed that $\Delta_\ell$ is larger than unity when $\mu$ is very large. Then, the radiative correction exceeds the tree level contribution to the lepton mass. For $\Delta_\ell \gg 1$, the Yukawa coupling constant is suppressed by $\Delta_\ell$, as observed by Eq.~\eqref{eq:LRmixing}. This may be disfavored by a naturalness argument~\cite{Crivellin:2010gw}. In Fig.~\ref{fig:gminus2_stausmubound}, $\Delta_\mu$ becomes larger than unity above the black dashed line, and above the blue dotted line in Fig.~\ref{fig:LRmixing}.


\section{Searches} \label{sec:search}

\begin{table}[tb]
\begin{center}
  \begin{tabular}{l|cccc} \hline
    Mass spectrum & Smuon & Vacuum  & LHC/ILC & LFV/EDM \\ \hline \hline
    $m_{\tilde e} = m_{\tilde\mu} = m_{\tilde\tau}$ (Sec.~\ref{sec:universal}) & $< 330/460\GeV$ & $\tilde\tau$ & \checkmark &  \\  
    $m_{\tilde e} = m_{\tilde\mu} < m_{\tilde\tau}$ (Sec.~\ref{sec:non-univ}) & $< 1.4/1.9\TeV$  & $\tilde\tau$ or $\tilde\mu$  &  & \checkmark \\ \hline
  \end{tabular}
  \caption{Summary of searches. The lightest smuon mass is restricted to explain the muon $g-2$ discrepancy at the $1\sigma/2\sigma$ level. The left-right mixing is limited by the vacuum stability condition of the stau-- or smuon--Higgs potential, depending on the stau masses. Models of the universal mass spectrum can be tested by LHC/ILC, while those of the non-universal spectrum predict large LFV/EDM.}
  \label{tab:summary}
\end{center}
\end{table}

In the previous section, it has been shown that when the muon $g-2$ anomaly \eqref{eq:g-2_deviation} is solved by the Bino--smuon contribution \eqref{eq:gminus2neut}, soft masses of the Bino and the smuon are bounded from above by the vacuum stability condition \eqref{eq:kyb}. The result is summarized in Tab.~\ref{tab:summary}. In this section, we study experimental status and future prospects to search for such SUSY models. The mass bounds depend on the slepton mass spectrum. When the stau is degenerate with the smuon, a tight constraint is imposed by the stau stability condition. In Sec.~\ref{sec:universal}, it will be discussed that the limit is strong enough for the superparticles to be detectable directly in colliders. On the other hand, if the staus are heavier or decoupled, the stability bound is relaxed. The superparticle masses can exceed the collider sensitivities. In Sec.~\ref{sec:non-univ}, it will be argued that such hierarchical mass spectrum can be probed by LFV and CPV.

\subsection{Universal slepton mass} \label{sec:universal}

In this section, we study collider searches for the SUSY models with the universal slepton mass spectrum. Here, the left- (right-) handed selectron, smuon and stau have a common soft SUSY-breaking mass,
\begin{align}
m_{\tilde e_L} = m_{\tilde\mu_L} = m_{\tilde\tau_L},~~~
m_{\tilde e_R} = m_{\tilde\mu_R} = m_{\tilde\tau_R}.
\end{align}
Then, the vacuum stability condition from the stau--Higgs potential restricts slepton masses up to $330-460\GeV$ to solve the muon $g-2$ anomaly. This is within the reach of the LHC or ILC sensitivity. In fact, some of the parameter regions have already been excluded by LHC, as shown later. 

Collider signatures depend on the lightest superparticle (LSP). In the case of universal soft slepton masses, either the lightest neutralino or the lightest stau is LSP among the MSSM particles.\footnote{
The lightest stau is lighter than sneutrinos when $\mu$ is large. 
}
In the latter case, the stau is likely to be long-lived.\footnote{
The stau could decay in detectors, for instance, through R-parity violations. The SUSY signatures depend on decay channels. Such cases are not discussed here. 
}
The (meta-) stable staus leave charged tracks in detectors. Such signatures have been studied by ATLAS~\cite{ATLAS2013058} and CMS~\cite{Chatrchyan:2013oca}. The CMS constraint on the cross section of the stau direct production provides the 95\% CL exclusion limit, $m_{\tilde\tau_1} > 339\GeV$. If this is imposed in addition to the vacuum stability condition, all the parameter regions of the long-lived stau are excluded in Fig.~\ref{fig:MSL-M1}. 

\begin{figure}[tbp]
 \begin{center}
 \includegraphics[width=7cm]{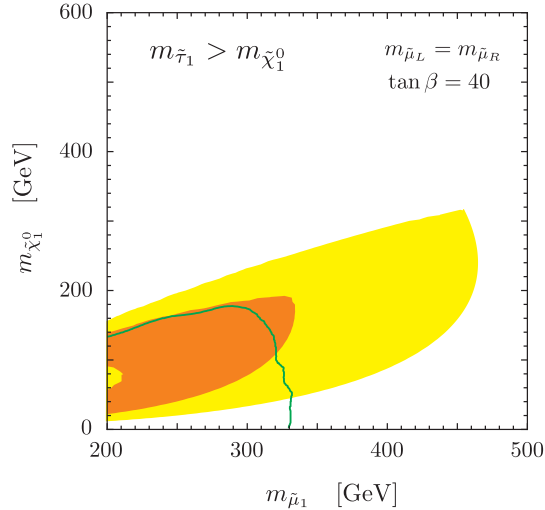}
 \caption{Same as Fig.~\ref{fig:MSL-M1}, but with a condition that the neutralino is LSP, $m_{\tilde\tau_1} > m_{\tilde\chi^0_1}$.}
 \label{fig:M1smu_plane}
 \end{center}
 \end{figure}
 
\begin{figure}[tbp]
 \begin{center}
 \includegraphics[width=7cm]{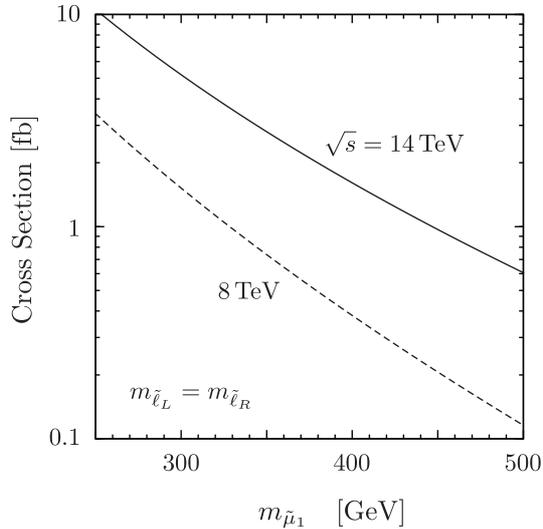}
 \caption{The total cross section of the smuon productions at LHC with $\sqrt{s} = 8\TeV$ (dashed) and $14\TeV$ (solid). The parameters satisfy $m_{\tilde\ell_L} = m_{\tilde\ell_R} = 2M_1$. The left-right mixing is maximized under the vacuum stability condition with $\tan\beta=40$.}
 \label{fig:LHC_smuon}
 \end{center}
 \end{figure}

In Fig.~\ref{fig:M1smu_plane}, we show the SUSY contributions to the muon $g-2$ when the left-right mixing is maximized under the two conditions; (i) the vacuum stability and (ii) the neutralino LSP, $m_{\tilde\tau_1} > m_{\tilde\chi^0_1}$. Then, SUSY signature is (opposite-sign same-flavor) di-lepton with large missing transverse energy. Selectrons and smuons are produced by collisions. They subsequently decay into the lightest neutralino and a partner lepton. Recently, this signature was studied by ATLAS~\cite{ATLAS2013049} and CMS~\cite{CMSPASSUS13006}. In particular, the 95\% CL exclusion limit is obtained for $m_{\tilde{\mu}_L} = m_{\tilde{\mu}_R}$ by ATLAS. The result is shown by the green solid line in Fig.~\ref{fig:M1smu_plane} (and in Fig.~\ref{fig:MSL-M1}). Here, the region below the line is excluded. In detail, the left-right mixing is negligible in the ATLAS analysis. However, the left- and right-handed smuons maximally mix with each other in Fig.~\ref{fig:M1smu_plane}. The total cross section of the smuon productions decreases by 10\% compared to the ATLAS setup. Since it is sufficiently small, the exclusion limit is considered to be almost the same as the ATLAS result. As a result, it is found that a large fraction of the $1\sigma$ parameter region of the muon $g-2$ is already excluded.\footnote{
In the ATLAS analysis, $m_{\tilde{e}} = m_{\tilde{\mu}}$ is assumed. The selectron production provides almost the same constraint as the smuon~\cite{ATLAS2013049}. Thus, if selectrons are decoupled, the mass bound becomes weaker. However, LFV/CPV constraints are severe, as will be mentioned in Sec.~\ref{sec:non-univ}.
}
The sensitivity will be improved by the upgrade of the energy and the luminosity. 
In Fig.~\ref{fig:LHC_smuon}, the total cross section of the smuon productions are presented for $\sqrt{s}=8\TeV$ and $14\TeV$. The cross section is estimated at the leading order.
For instance, it becomes $1\,{\rm fb}$ for the lightest smuon mass of $330\GeV$ at $\sqrt{s}=8\TeV$, which corresponds to the current LHC bound. The same cross section is obtained for $450\GeV$ at $14\TeV$.
Studies for the future sensitivity are required. 

When the neutralino is LSP, the stau productions have also been studied in LEP and LHC. If their mass difference is larger than $15\GeV$, the stau mass is constrained to be larger than $81.9\GeV$ by LEP~\cite{PDG}. LHC is still ineffective to search for di-tau events from the direct stau productions~\cite{ATLAS2013028}. These limits are sufficiently weak. Fig.~\ref{fig:M1smu_plane} does not change even if they are imposed. 

\begin{table}[tbp]
\begin{center}
\begin{tabular}{|c|ccc|ccccccc|c|}
\hline
&
$m_{\tilde\ell}$ & $M_1$ & $\mu$ & 
$m_{\tilde{e}_1}$ & $m_{\tilde{e}_2}$ & 
$m_{\tilde{\mu}_1}$ & $m_{\tilde{\mu}_2}$ & 
$m_{\tilde{\tau}_1}$ & $m_{\tilde{\tau}_2}$ & 
$m_{\tilde{\chi}^0_1}$ 
& $\Delta a_\mu$ 
\\ \hline
A &
300 & 200 & 756 &
303 & 304 &
298 & 309 &
199 & 380 &
199 & 16.1
\\
$\rm{A^{\prime}}$&
300 & 200 & 699 &
303 & 304 &
299 & 308 &
209 & 375 &
199 & 14.6
\\
B &
470 & 250 & 1680 &
472 & 472 &
465 & 479 &
329 & 581 &
250 & 10.2
\\
C &
340 & 160 & 1138 &
343 & 343 &
336 & 350 &
199 & 442 &
160 & 18.0
\\ \hline
\end{tabular}
\caption{Model parameters and mass spectra at several model points in Fig.~\ref{fig:M1smu_plane}. The masses are in units of GeV, and the muon $g-2$ is scaled by $10^{-10}$. Here, $m_{\tilde\ell}$ denotes $m_{\tilde\ell_L} = m_{\tilde\ell_R}$, and $\tan\beta=40$ is set.}
\label{tab:mass}
\end{center}
\end{table} 

In Tab.~\ref{tab:mass}, superparticle mass spectra are listed for several points in Fig.~\ref{fig:M1smu_plane}. The lightest neutralino mass is close to the Bino mass, since $\mu$ is very large. Given $m_{\tilde\ell} \equiv m_{\tilde\ell_L} = m_{\tilde\ell_R}$, the slepton masses are hierarchical except for those of the selectrons due to a large left-right mixing. The lightest stau mass is closest to the neutralino mass among the sleptons. In Fig.~\ref{fig:M1smu_plane}, the point A (A') is around the upper side of the contour of the muon $g-2$. In the vicinity of this side, $m_{\tilde\tau_1}$ has the closest value to $m_{\tilde\chi^0_1}$ in the allowed range. If $m_{\tilde\tau_1} > m_{\tilde\chi^0_1}$ is imposed, it satisfies $m_{\tilde\tau_1} = m_{\tilde\chi^0_1}$ to maximize the SUSY contributions to the muon $g-2$ around the upper side of the contour. Above it, the regions are excluded by the long-lived stau search, or the muon $g-2$ becomes too small. On the other hand, if the mass difference between the stau and the neutralino is assumed to be larger than $\delta M$, they satisfy $m_{\tilde\tau_1}=m_{\tilde\chi^0_1}+\delta M$  in the vicinity of the upper side. At the point A', $\delta M = 10\GeV$ is imposed, and $m_{\tilde\tau_1}$ is larger than $m_{\tilde\chi^0_1}$ by $10\GeV$. In contrast, the points B and C are away from this region. B (C) is close to the maximal end point of the lightest smuon mass which explains the muon $g-2$ at the $2\sigma$ ($1\sigma$) level. Here, the left-right mixing is determined by the vacuum stability condition of the stau. 

\begin{figure}[tbp]
 \begin{center}
 \includegraphics[width=7cm]{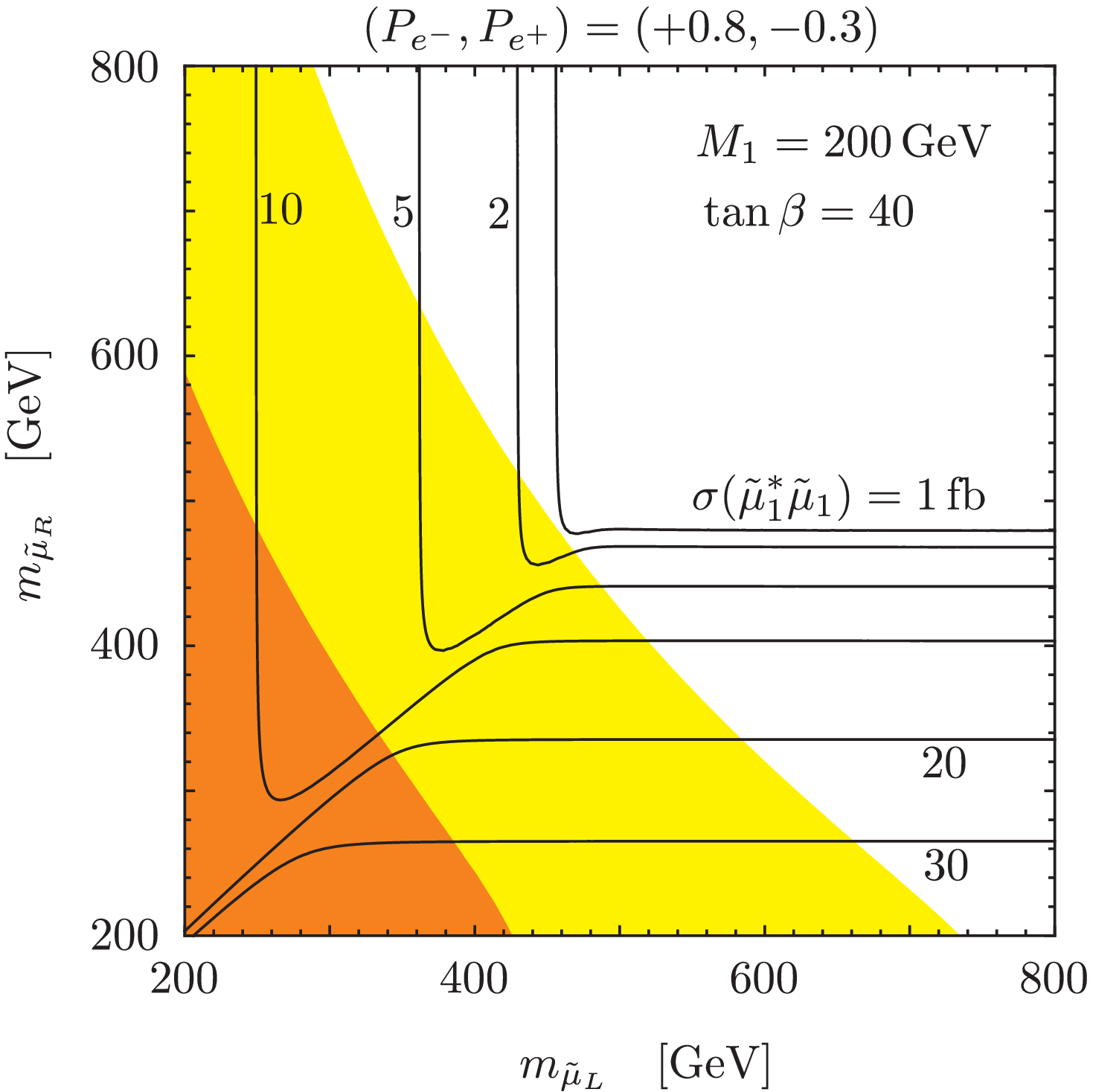}\hspace*{5mm}
 \includegraphics[width=7cm]{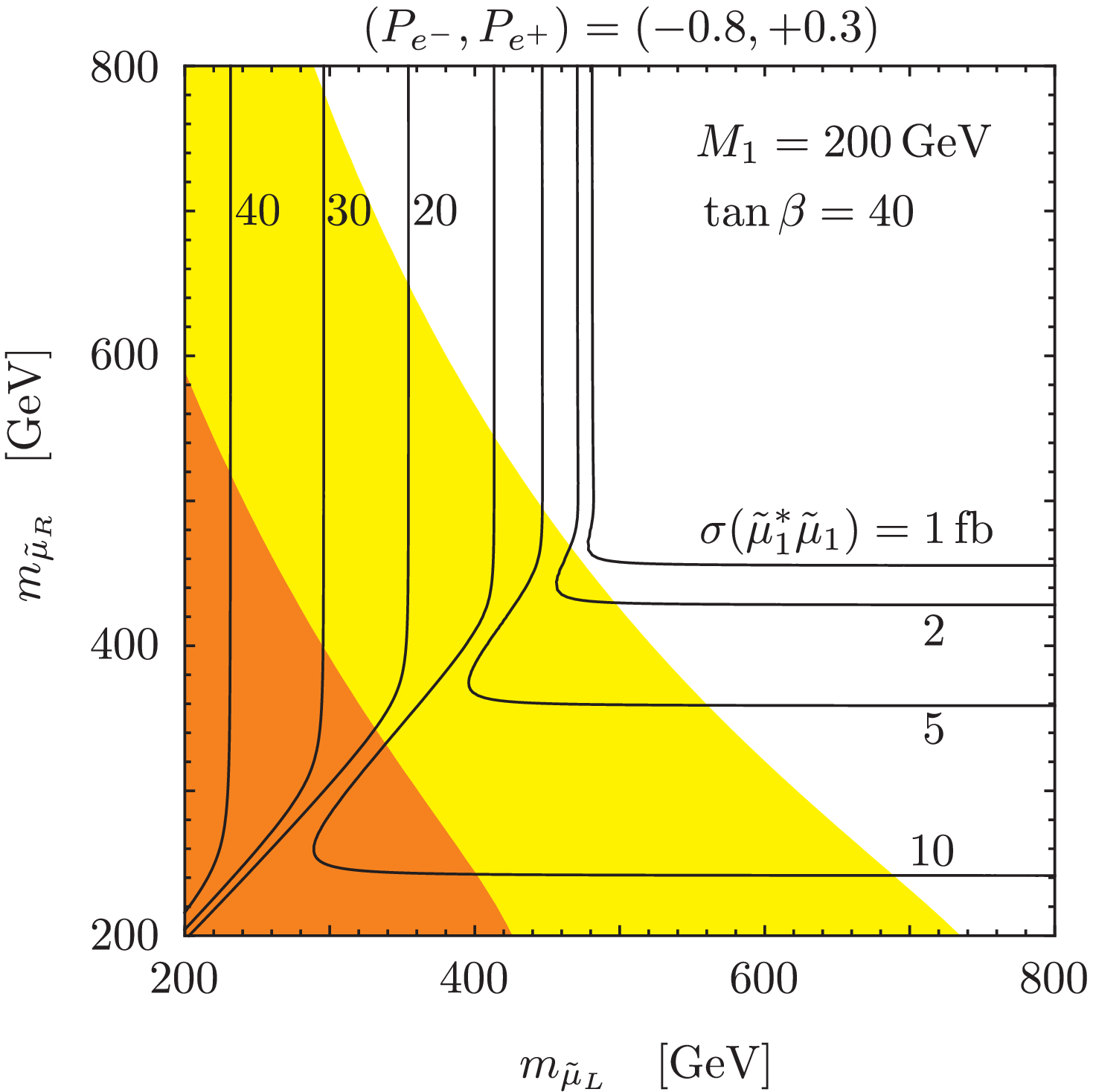}
  \end{center}
 \caption{The production cross section of the lightest smuon at ILC with $\sqrt{s} = 1\TeV$. The beams are assumed to be polarized at 80\% $(e^-)$ and 30\% $(e^+)$. The left-right mixing is maximized under the vacuum stability condition with $\tan\beta=40$. The $1\sigma$ ($2\sigma$) regions of the muon $g-2$ are also shown by the orange (yellow) bands. The parameters are $M_1=200\GeV$ and $M_{\rm soft} = 10\TeV$.} 
  \label{fig:SmuonProdILC}
\end{figure}

Linear colliders of $e^+e^-$ (ILC) are very useful (see e.g., Ref.~\cite{Baer:2013cma,Baer:2013vqa}). They can provide rich informations of the models. Moreover, they are superior to LHC when the slepton masses are close to that of the lightest neutralino~\cite{Martyn:2004jc}. In the orange/yellow regions of Fig.~\ref{fig:M1smu_plane}, selectrons, smuons and staus can be produced at the linear colliders. In Fig.~\ref{fig:SmuonProdILC}, production cross sections of the lightest smuon are shown for the left- and right-handed smuon masses. Here, it is assumed that the collision energy is $\sqrt{s} = 1\TeV$ at ILC, and the beams are polarized at 80\% for the electron and 30\% for the positron. The left-right mixing is maximized under the vacuum stability condition. The other parameters are $M_1=200\GeV$ and $\tan\beta=40$. It is found that the cross section is larger than 20 (1)$\invfb$ for the $1\sigma$ ($2\sigma$) parameter region of the muon $g-2$. The estimation is based on Ref.~\cite{Freitas:2003yp}.\footnote{
The cross section is calculated at the leading order in Fig.~\ref{fig:SmuonProdILC}. It can be enhanced by $\sim 10\%$ near the mass threshold~\cite{Freitas:2003yp}. 
} 
It is expected that the smuons can be discovered in almost all the parameter region that are kinematically allowed (cf.~\cite{Baer:2013vqa}).

It is possible to measure masses of the smuon and the neutralino from event distributions and the cross section~\cite{Tsukamoto:1993gt,Nojiri:1996fp}. Also, the chirality structure of the smuons could be determined by the beam polarization. We assume $(P_{e^-},P_{e^+}) = (+0.8,-0.3)$ in the left panel of Fig.~\ref{fig:SmuonProdILC}, and $(-0.8,+0.3)$ in the right panel. The cross section of a chiral smuon is sensitive to the polarization, because the productions proceed by the s-channel $\gamma/Z$ exchanges. 
On the other hand, when $m_{\tilde\mu_L}=m_{\tilde\mu_R}$, the cross section is insensitive to the beam polarization. Since the left-right mixing is large, the left- and right-handed smuons are maximally mixed with each other. Both of them can be produced, and the mass difference is expected to be measured, for instance, by measuring threshold productions. At the model points in Tab.~\ref{tab:mass}, the difference is $\gtrsim 10\GeV$, which is much larger than the uncertainty of the mass measurement at ILC, $\Order(10-100)\MeV$~\cite{Baer:2013cma}. It is emphasized that the smuon productions are clean and direct signatures of the SUSY contributions to the muon $g-2$. 

\begin{figure}[tbp]
 \begin{center}
 \includegraphics[width=7cm]{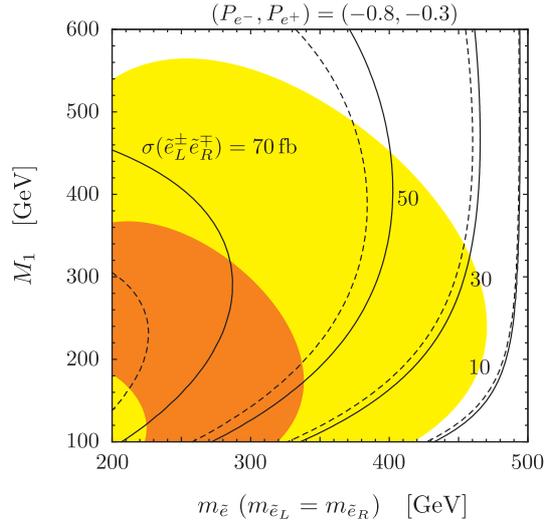}
  \end{center}
 \caption{The production cross section of $\tilde e^+_R \tilde e^-_L$ and $\tilde e^+_L \tilde e^-_R$ at ILC with $\sqrt{s} = 1\TeV$. The parameters are same as Fig.~\ref{fig:SmuonProdILC}, but $M_1$ is varied with $m_{\tilde e_L}=m_{\tilde e_R} \equiv m_{\tilde e}$. The corrections to the Bino coupling due to decoupling of heavy superparticles are included (discarded) in the black solid (dashed) lines.} 
  \label{fig:SelectronProdILC}
\end{figure}

The selectron productions proceed not only by the s-channel $\gamma/Z$ exchanges, but also by the t-channel Bino exchange. The latter contribution can enhance the cross section and provide individual information in addition to mass measurements of the selectron and the neutralino. The productions, $e^+_L e^-_L \to \tilde e^+_R \tilde e^-_L$ and $e^+_R e^-_R \to \tilde e^+_L \tilde e^-_R$, proceed by the t-channel. In Fig.~\ref{fig:SelectronProdILC}, their total cross sections are shown by the black solid lines. Here, the parameters are $m_{\tilde e_L} = m_{\tilde e_R}$, and the beam polarization is 80\% for the electron and 30\% for the positron. It is found that the cross sections are $\Order(10)\,{\rm fb}$ in the muon $g-2$ parameter regions. They are larger than those of the smuon in Fig.~\ref{fig:SmuonProdILC}. Importantly, this channel is useful to measure the Bino--electron--selectron couplings~\cite{Nojiri:1996fp,Nojiri:1997ma}. They are deviated from the U(1)$_Y$ gauge coupling constant by decoupling heavy superparticles, as discussed in Sec.~\ref{sec:g-2}. In Fig.~\ref{fig:SelectronProdILC}, the cross sections are estimated with (without) $\delta\tilde g_L$ and $\delta\tilde g_R$~(see Eqs. (\ref{eq:gL}) and (\ref{eq:gR})). The results are shown by the solid (dashed) lines. It is found that 
the corrections enhance the cross section by $8-10\%$ for $M_{\rm susy} = 10\TeV$. This is measurable at ILC~\cite{Nojiri:1996fp,Nojiri:1997ma}. 

Cross sections of $e^+_L e^-_R \to \tilde e^+_R \tilde e^-_R$ and $e^+_R e^-_L \to \tilde e^+_L \tilde e^-_L$ also include the t-channel contribution of the Bino (see e.g., Ref.~\cite{Freitas:2003yp}). In particular, the former cross section is enhanced well compared to those solely by the s-channel $\gamma/Z$ exchanges. It can be $\gtrsim 100\,{\rm fb}$ when $M_1$ is relatively small in the muon $g-2$ parameter regions. However, in the latter process, the t-channel contribution interferes destructively with the s-channel contribution. On the other hand, the cross section of $e^+_L e^-_R \to \tilde e^+_R \tilde e^-_R$ differs by $5-6\%$ between the cases with and without $\delta\tilde g_L$ and $\delta\tilde g_R$. This is smaller than the above channels. 

The stau productions proceed similarly to the smuon at ILC. The stau mass parameters can be determined~\cite{Nojiri:1994it,Nojiri:1996fp,Boos:2003vf}. The produced staus decay into the lightest (Bino-like) neutralino and a tau lepton. The tau polarization can be measured from the energy spectra of tau hadronic decays~\cite{Nojiri:1994it}. If productions of both the heavy and light staus are kinematically allowed, all the components of the stau mass matrix can be measured. In particular, the left-right mixing is determined.\footnote{
Strictly, it is difficult to determine $\mu$ and $A_\tau$ separately by the stau productions, because the lightest neutralino is almost composed of the Bino~\cite{Nojiri:1996fp,Boos:2003vf}.
}

Let us discuss the dark matter in the present model. The lightest neutralino is a candidate of the dark matter. The neutralino relic abundance becomes consistent with the measured cold dark matter abundance~\cite{Hinshaw:2012aka,Ade:2013zuv} by the stau co-annihilation. The mass difference between the stau and the neutralino is required to be $5-10\GeV$. This corresponds to the model points in the vicinity of the upper side of the muon $g-2$ contours in Fig.~\ref{fig:M1smu_plane}, including the points A or A'. This region has not been excluded by the studies of the di-lepton signature at LHC~\cite{ATLAS2013049,CMSPASSUS13006}. Masses of the selectrons and the smuons are too close to that of the neutralino for LHC. 
ILC is superior to study the region~\cite{Baer:2013cma,Baer:2013vqa}. Selectrons and smuons as well as staus can be produced, as discussed above. 

Finally, let us mention the branching ratios of the Higgs boson decays. When the staus are light, and their left-right mixing is large, the branching ratio of the di-photon channel can be enhanced sizably~\cite{Carena:2011aa}. We have estimated it in the muon $g-2$ parameter regions of Fig.~\ref{fig:M1smu_plane}. In the $1\sigma$ parameter region, the ratio increases by $10-40\%$ compared to the SM prediction, while it is enhanced by $5-10\%$ in the $2\sigma$ region. The latter is comparable to the sensitivity of the high-luminosity LHC with $\int\mathcal{L}=3000\invfb$~\cite{ATLAS2012001,CMS:2013xfa}. On the other hand, the ratio of the Higgs boson decaying to the di-muon is almost unchanged. Even when the radiative corrections to the Higgs boson coupling with the muon is very large, i.e., $\Delta_\mu \gg 1$, the ratio does not change, because the tree level (Yukawa) coupling decreases by $\Delta_\mu$, and the sum of the tree and radiative contributions is not varied.

\subsection{Non-universal slepton mass} \label{sec:non-univ}

\begin{figure}[tbp]
 \begin{center}
 \includegraphics[width=7cm]{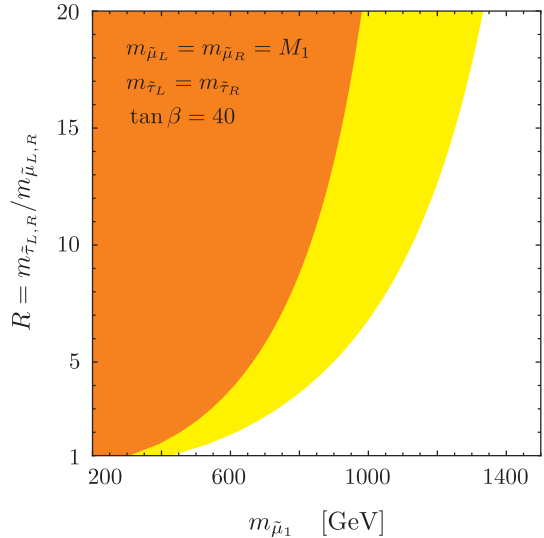}
 \caption{The lower bound on the stau mass to satisfy the vacuum stability bound of the stau--Higgs potential. In the orange (yellow) region, the SUSY contribution to the muon $g-2$ can explain the discrepancy \eqref{eq:g-2_deviation} at the $1\sigma$ $(2\sigma)$ level. The parameters are $m_{\tilde{\mu}_L} = m_{\tilde{\mu}_R}= M_1$, $m_{\tilde{\tau}_L} = m_{\tilde{\tau}_R}$, $\tan\beta = 40$ and $M_{\rm soft} = 30\TeV$. }
  \label{fig:boundR}
 \end{center}
 \end{figure}

In this section, we discuss the case without slepton mass universality. The slepton soft SUSY-breaking masses satisfy 
\begin{align}
m_{\tilde e_L} = m_{\tilde\mu_L} < m_{\tilde\tau_L},~~~
m_{\tilde e_R} = m_{\tilde\mu_R} < m_{\tilde\tau_R}.
\label{eq:non-univ}
\end{align}
As discussed in Sec.~\ref{sec:vac}, the vacuum stability bound of the stau--Higgs potential is relaxed by heavy staus, and the smuon masses are allowed to be larger in order to solve the muon $g-2$ anomaly. In other words, the stau masses are required to be large relative to the smuons, when the smuons are heavy. In Fig.~\ref{fig:boundR}, the lower bound on the stau mass is shown. In the orange (yellow) regions, the muon $g-2$ anomaly is solved at the $1\sigma$ ($2\sigma$) level, while the vacuum stability constraint is avoided. The ratios of the stau and smuon masses are defined as
\begin{align}
R_L \equiv \frac{m_{\tilde{\tau}_L}}{m_{\tilde{\mu}_L}},~~~
R_R \equiv \frac{m_{\tilde{\tau}_R}}{m_{\tilde{\mu}_R}}.
\end{align}
The parameters are $M_1 = m_{\tilde{\mu}_L} = m_{\tilde{\mu}_R}$, $m_{\tilde{\tau}_L} = m_{\tilde{\tau}_R}$, $\tan \beta = 40$ and $M_{\rm soft} = 30\TeV$. The vertical axis is $R \equiv R_L = R_R$. If the staus are decoupled, the left-right mixing is bounded either by Eq.~\eqref{eq:mLRmax} or the vacuum stability bound of the smuon--Higgs potential, as discussed in Sec.~\ref{sec:vac}. The smuon masses can be $1.4\TeV$ $(1.9\TeV)$ for the $1\sigma$ $(2\sigma)$ level of the muon $g-2$. Even though it becomes difficult to produce such sleptons at LHC or ILC, non-universal slepton masses generically cause problems of too large lepton flavor violations (LFV) and CP violations (CPV).

LFV and CPV are sensitive to off-diagonal (generation mixing) components of the slepton mass matrices. They are suppressed if the slepton mass matrices are universal among the flavors, such as in the gauge mediated SUSY-breaking models. 
However, even when the mass matrices are diagonal in the model basis, sizable FCNC and CPV are generically induced as long as the diagonal components are not equal to each others.
In fact, a lot of models have been proposed to explain the SM Yukawa couplings, and many of them predict non-diagonal Yukawa matrices in the model basis. Even if the sfermion mass matrices are diagonal in this basis, off-diagonal components are generated in the fermion mass eigenstate basis, which are obtained by rotating the mass matrices of the model basis with unitary matrices. In other words, the super GIM mechanism does not work generically unless the sfermion mass matrices are universal. 

When the muon $g-2$ anomaly is solved by the SUSY contributions, LFV and lepton electric dipole moments (EDM) are generically sizable. In this section, the following setup is considered. In the model basis, the slepton mass matrices are diagonal among the flavors, 
\begin{align}
(m_{\tilde\ell_L}^2)_{ij} &= \text{diag}(m_{\tilde e_L}^2, m_{\tilde\mu_L}^2, m_{\tilde\tau_L}^2), \\
(m_{\tilde\ell_R}^2)_{ij} &= \text{diag}(m_{\tilde e_R}^2, m_{\tilde\mu_R}^2, m_{\tilde\tau_R}^2). 
\end{align}
Moreover, $m_{\tilde e_L} = m_{\tilde\mu_L}$ and $m_{\tilde e_R} = m_{\tilde\mu_R}$ are imposed. Otherwise, LFV becomes too large, as shown later. The Yukawa matrices are generally non-diagonal in the model basis. The mass eigenstate basis of the charged leptons is obtained by the left- and right-handed unitary matrices, $U_L, U_R$, as
\begin{align}
U_R\, M_\ell\, U_L^\dagger \equiv U_R\, Y_\ell\, v_d\, U_L^\dagger
={\rm diag}
\left( \frac{m_e}{1+\Delta_e}, ~\frac{m_\mu}{1+\Delta_\mu}, ~\frac{m_\tau}{1+\Delta_\tau} \right),
\end{align}
where $v_d$ is the down-type Higgs VEV, and $\Delta_\ell$ is given by Eq.~\eqref{eq:Yukawa}. The unitary matrices are generally represented as
\begin{equation}
U_{L,R} = \exp 
\left[
\begin{pmatrix}
 0 & (\delta_{L,R})_{12} & (\delta_{L,R})_{13} \\
 -(\delta_{L,R})^{* }_{12} & 0 & (\delta_{L,R})_{23} \\
-(\delta_{L,R})^{* }_{13}  & -(\delta_{L,R})^{*}_{23} & 0 \\
\end{pmatrix} 
\right].
\label{eq:unitary}
\end{equation}
Non-vanishing mixings, $(\delta_{L,R})_{ij}$, induce LFV and EDM, as long as the slepton masses are non-universal.\footnote{
Similar setup has been studied in Ref.~\cite{Endo:2010fk}, where squarks of the first two generations are light.
}
Note that quark FCNC and CPV are suppressed by heavy colored superparticles in this paper.

The magnetic dipole contributions to LFV are represented by the following effective Lagrangian,
\begin{align}
\mathcal{L}_{\rm eff} =
e \frac{m_{\ell_j}}{2}
\bar\ell_i \sigma_{\mu\nu} \left( A^L_{ij} P_L + A^R_{ij} P_R \right) \ell_j F^{\mu\nu} + {\rm h.c.},
\label{eq:L-LFV}
\end{align}
where $i, j$ are flavor indices. Contributions to the other higher dimensional operators are subdominant in this paper. The Wilson coefficients, $A^L_{ij}$ and $A^R_{ij}$, are dominated by the Bino--slepton contributions, similarly to the muon $g-2$. In the mass insertion approximation, they are estimated as (cf.~\cite{Cho:2001hx})
\begin{align}
A^L_{ij} &= (1 + \delta^{\rm 2loop}) \frac{\alpha_Y}{8\pi}
\frac{M_1\mu\tan\beta}{m_{\ell_j}} \sum_{a,b=1, 2, 3} 
\Big[ U_R \Big]_{ib} \Big[ M_\ell \Big]_{ba} \Big[ U_L^\dagger \Big]_{aj}\,
F_{a,b}, 
\label{eq:LFVaml} \\ 
A^R_{ij} &= (1 + \delta^{\rm 2loop}) \frac{\alpha_Y}{8\pi}
\frac{M_1\mu\tan\beta}{m_{\ell_j}} \sum_{a,b=1, 2, 3}
\Big[ U_L \Big]_{ia} \Big[ M_\ell^\dagger \Big]_{ab} \Big[ U_R^\dagger \Big]_{bj}\,
F_{a,b}.
\label{eq:LFVamr} 
\end{align}
The two loop factor $\delta^{\rm 2loop}$ is found in Eq.~\eqref{eq:2-loop}. The loop function $F_{a,b}$ is defined as
\begin{align}
F_{a,b} = 
\frac{1}{m_{\tilde\ell_{La}}^2 m_{\tilde\ell_{Rb}}^2} 
f_N \left( \frac{m_{\tilde\ell_{La}}^2}{|M_1|^2}, \frac{m_{\tilde\ell_{Rb}}^2}{|M_1|^2} \right).
\end{align}
The muon FCNC's are the most sensitive to the non-universal slepton mass matrices and the non-diagonal Yukawa matrix. The decay rate of $\mu \to e\gamma$ is represented as 
\begin{align}
\Gamma(\mu \to e\gamma) 
= \frac{\alpha}{4} m_\mu^5 \left( \left|A^L_{12}\right|^2 + \left|A^R_{12}\right|^2 \right),  \label{eq:LFV}
\end{align}
where the unitarity matrices in $A^R_{12}$ are expanded as
\begin{align}
\sum_{a,b=1, 2, 3} 
\Big[ U_L \Big]_{1a} \Big[ M_\ell^\dagger \Big]_{ab} \Big[ U_R^\dagger \Big]_{b2}\, F_{a,b}
&=
- \frac{m_\mu}{1+\Delta_\mu} (\delta_L)_{12} \left(F_{1,2} - F_{2,2}\right) 
\\ 
&~~~ +\frac{m_\tau}{1+\Delta_\tau} (\delta_L)_{13} (\delta_R)_{23}^* 
\left(F_{1,2} - F_{1,3} - F_{3,2} + F_{3,3}\right), 
\notag 
\end{align}
at the leading order of $(\delta_L)_{ij}$, $(\delta_R)_{ij}$ and $m_\mu/m_\tau$. Here and hereafter, $m_e = 0$ is set, for simplicity. Similarly, $A^L_{12}$ is obtained by replacing $L \leftrightarrow R$. In the last term, $F_{1,2}$ is dominant when the staus are heavy, i.e., $F_{1,2} \gg F_{1,3}, F_{3,2}, F_{3,3}$ for $m_{\tilde e_L}, m_{\tilde\mu_L} \ll m_{\tilde\tau_L}$ and $m_{\tilde e_R}, m_{\tilde\mu_R} \ll m_{\tilde\tau_R}$. On the other hand, the right-hand side vanishes when the slepton masses are universal, as expected from the super GIM mechanism. In particular, when the sleptons are degenerate in the first two generations, the first term becomes zero due to $F_{1,2}=F_{2,2}$. Otherwise, the muon LFV is induced at the order of $(\delta_L)_{12}$.

The above rate is compared with the SUSY contribution to the muon $g-2$. In the non-universal slepton mass spectrum, it is represented by $A^L_{ij}$ and $A^R_{ij}$ as
\begin{align}
a_\mu({\rm SUSY})
&= m_\mu^2\, {\rm Re} \left[ A^L_{22} + A^R_{22} \right] \notag \\
&= (1 + \delta^{\rm 2loop}) \frac{\alpha_Y}{4\pi} 
m_\mu M_1\mu\tan\beta 
\left[ 
\frac{m_\mu}{1+\Delta_\mu} F_{2,2} + \kappa
\right]. 
\label{eq:gmin2_non-univ}
\end{align}
This is same as Eq.~\eqref{eq:gminus2neut} up to a correction $\kappa$, which is represented as
\begin{align}
\kappa = \frac{m_\tau}{1+\Delta_\tau} {\rm Re} \left[ (\delta_L)_{23} (\delta_R)_{23}^* \right]
\left(F_{2,2} - F_{2,3} - F_{3,2} + F_{3,3}\right) + \cdots. 
\end{align} 
Here, the omitted terms are suppressed by $(\delta_L)_{ij}$, $(\delta_R)_{ij}$ or $m_\mu/m_\tau$. 
If the slepton mass matrices are universal, $\kappa $ vanishes.  

It is noticed that Eqs.~\eqref{eq:LFV} and \eqref{eq:gmin2_non-univ} are tightly correlated to each other. For the mass spectrum \eqref{eq:non-univ} with $R_L, R_R \gg 1$, the ratio is
\begin{align}
\frac{{\rm Br}(\mu \to e \gamma)}{a_\mu({\rm SUSY})^2} \simeq 
\frac{1}{\Gamma_{\rm tot}} \frac{\alpha m_{\mu}}{16}\,
\big|\delta_{13}\delta_{23}\big|^2
\left( \frac{m_{\tau}}{m_{\mu}} \frac{1 + \Delta_{\mu}}{1 + \Delta_{\tau}} \right)^2,
\label{eq:muegammag-2}
\end{align}
where $\Gamma_{\rm tot}$ is the total decay rate of muon, and the mixing is defined as
\begin{align}
\big|\delta_{13}\delta_{23}\big|^2 \equiv 
\big|(\delta_R)_{13}(\delta_L)_{23}\big|^2 + \big|(\delta_L)_{13}(\delta_R)_{23}\big|^2.
\label{eq:mixing}
\end{align}
It is independent of superparticle mass spectra except through $\Delta_{\mu}$ and $\Delta_{\tau}$. Thus, when the muon $g-2$ discrepancy \eqref{eq:g-2_deviation} is explained by the SUSY contributions, $\mu \to e\gamma$ is induced by the non-universal slepton mass sizably. It is important that the decay is not suppressed by heavy slepton masses, for instance, $m_{\tilde\mu_1} = 1.4\TeV$ or $1.9\TeV$ in Tab.~\ref{tab:summary}, for given SUSY contributions to the muon $g-2$. 

\begin{figure}[tb]
\begin{center}
 \includegraphics[width=7cm]{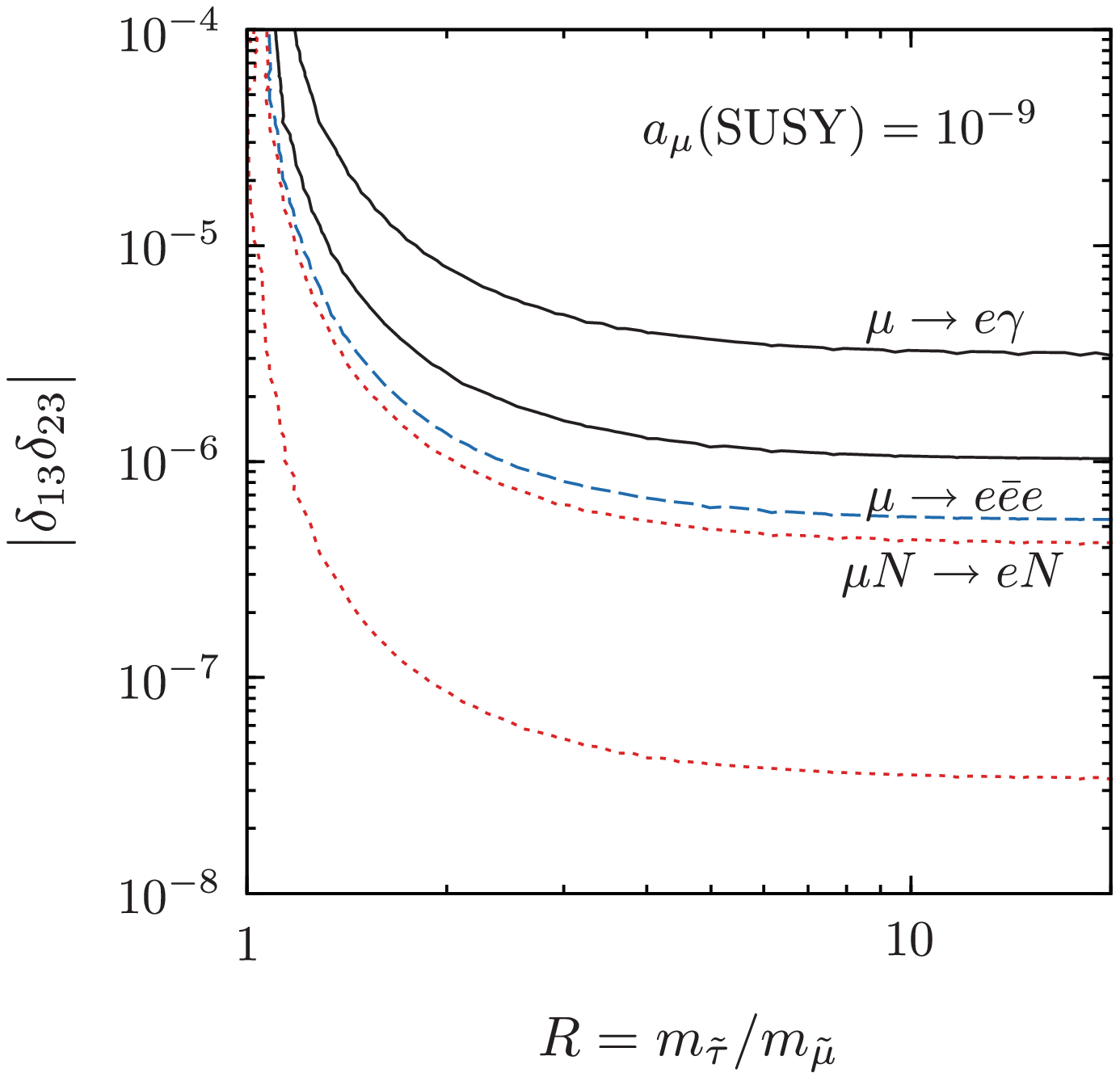}\hspace*{2mm}
 \includegraphics[width=7cm]{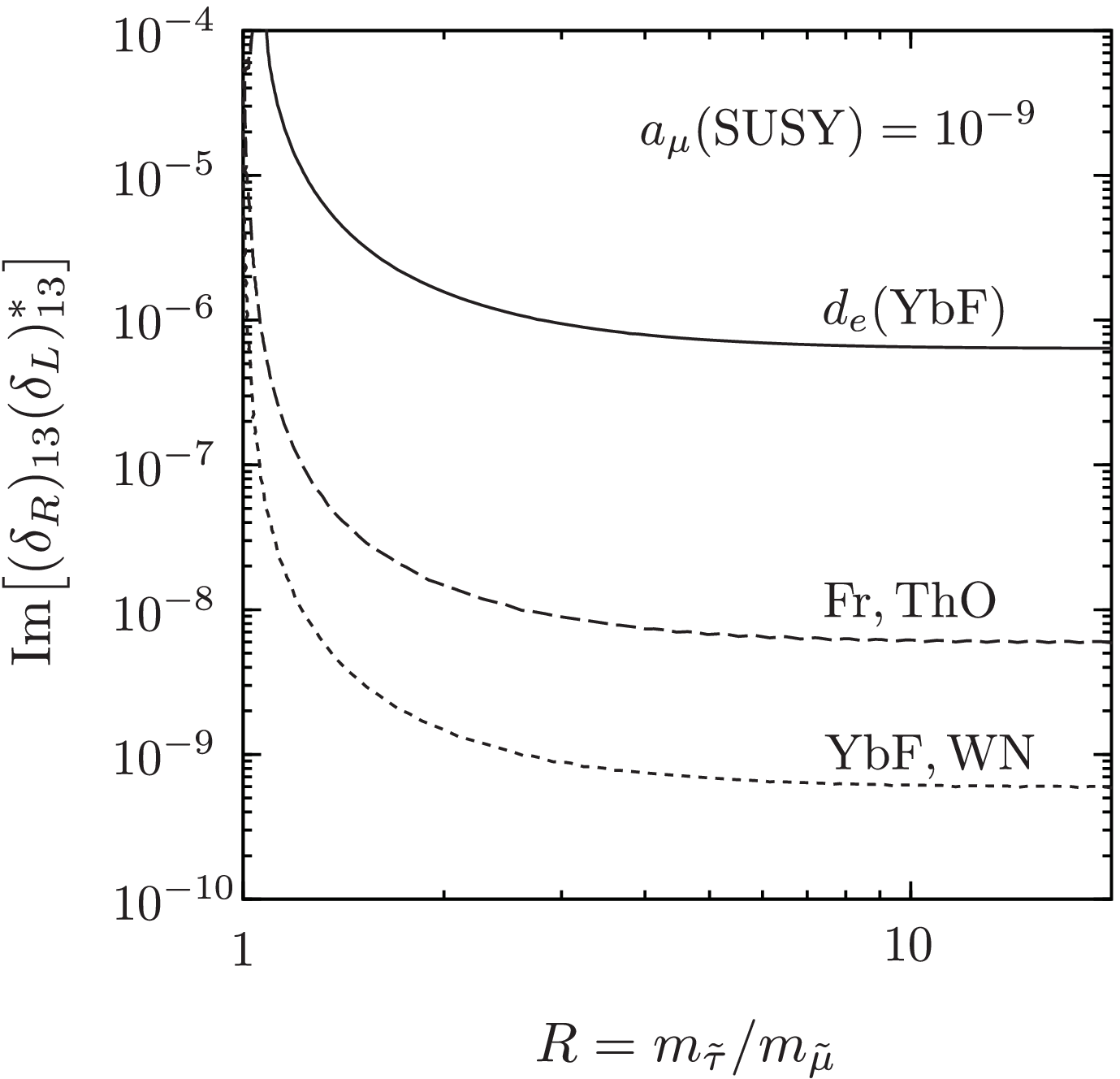}
 \caption{Contours of ${\rm Br}(\mu \to e \gamma)$ (left) and $d_e$ (right) for $a_\mu({\rm SUSY}) = 10^{-9}$ with $M_{\rm soft} = 30\TeV$. In the left panel, the upper black solid line is the current bound by MEG, ${\rm Br}(\mu \to e \gamma) < 5.7 \cdot 10^{-13}$ at 90\% CL~\cite{Adam:2013mnn}. The lower one is the sensitivity of the MEG upgrade, ${\rm Br}(\mu \to e \gamma) = 6 \cdot 10^{-14}$~\cite{Baldini:2013ke}. The blue dashed line is the Mu3e sensitivity, ${\rm Br}(\mu \to e\bar ee) = 10^{-16}$~\cite{Blondel:2013ia}. The $\mu-e$ conversion is expected to probe down to $R_{\mu e} = 3 \cdot 10^{-17}$ (upper red dotted) by COMET/Mu2e~\cite{Kuno:2013mha,Abrams:2012er} and $2 \cdot 10^{-19}$ (lower red dotted) by PRISM/PRIME~\cite{Kuno:2012pt}. 
 In the right panel, the black solid line is the current bound by using YbF, $|d_e| < 1.05 \cdot 10^{-27}\,e{\rm cm}$ at 90\% CL~\cite{Hudson:2011zz}. The sensitivity is planned to be improved: $|d_e| = 10^{-29}\,e{\rm cm}$ by Fr or ThO~\cite{Sakemi:2011zz,Vutha:2009ux,Campbell:2013ipa}, and $10^{-30}\,e{\rm cm}$ by YbF or WN~\cite{Kara:2012ay,Kawall:2011zz}.
 Here, $m_{\tilde\ell_L} = m_{\tilde\ell_R}$ is assumed.
}
 \label{fig:LFVCPV}
\end{center}
\end{figure}

In the left panel of Fig.~\ref{fig:LFVCPV}, contours of ${\rm Br}(\mu \to e \gamma)$ are shown. Here, $m_{\tilde\ell_L} = m_{\tilde\ell_R}$ is assumed. The SUSY contributions to the muon $g-2$ is fixed to be $a_\mu({\rm SUSY}) = 1 \times 10^{-9}$ with $M_{\rm soft} = 30\TeV$. The small corrections are taken at $M_1 = m_{\tilde{\mu}} = 400\GeV$ and $\tan\beta = 40$ as a reference, though the result is almost independent of them. In the figure, the contours correspond to the current limit and future sensitivities of experiments,
\begin{itemize}
\item the current limit of the MEG experiment, ${\rm Br}(\mu \to e \gamma) < 5.7 \times 10^{-13}$ at 90\% CL~\cite{Adam:2013mnn} (upper black solid line in the figure).
\item the sensitivity of the MEG upgrade, ${\rm Br}(\mu \to e \gamma) = 6 \times 10^{-14}$~\cite{Baldini:2013ke} (lower black solid line).
\item the sensitivity of the Mu3e experiment, ${\rm Br}(\mu \to e\bar ee) = 10^{-16}$ at Phase II~\cite{Blondel:2013ia} (blue dashed line).
\item the sensitivity of the COMET experiment, $R_{\mu e} = 3 \times 10^{-17}$ at Phase II~\cite{Kuno:2013mha} (upper red dotted line). The Mu2e experiment has a similar sensitivity~\cite{Abrams:2012er}.
\item the proposal of the PRISM/PRIME project, $R_{\mu e} = 2 \times 10^{-19}$~\cite{Kuno:2012pt} (lower red dotted line).
\end{itemize}
Note that $\mu \to e\bar ee$ and $\mu-e$ conversion experiments have better sensitivity in future than those of $\mu \to e \gamma$. In particular, the latter experiment has low (accidental) backgrounds. On the other hand, the current constraint and future sensitivities of the tau LFV's are weaker than those of the muon, though they are also induced by the non-universal slepton masses with finite $(\delta_L)_{23}$ and $(\delta_R)_{23}$. 

From the figure, it is found that the LFV decay rate increases rapidly as the staus become heavier than  the smuons, $R > 1$. When the staus are decoupled, $R \gg 1$, the mixing $|\delta_{13}\delta_{23}|$ in Eq.~\eqref{eq:mixing} is limited to be smaller than $3 \times 10^{-6}$ by MEG for $a_\mu({\rm SUSY}) = 1 \times 10^{-9}$. For instance, when the smuon and selectron masses are larger than $1\TeV$, the stau are required to be heavier than $7\TeV$ ($R > 7$) to avoid the vacuum stability constraint, according to Fig.~\ref{fig:boundR}. If the lepton Yukawa matrix is related to the quark sector, e.g., by the GUT relation, it is naively expected to be $|\delta_{13}\delta_{23}| \sim V_{ub} V_{cb} \sim 10^{-4}$. This already exceeds the above limit. Thus, non-universal slepton mass spectra are tightly constrained by LFV. In future, if sleptons are neither discovered at LHC nor ILC, the model is expected to be probed by LFV. Otherwise, the SM Yukawa matrices are tightly limited in the model basis, or when flavor models are constructed. 

The flavor off-diagonal components of the Yukawa matrices are sources of the CP violations, when the slepton mass matrices are non-universal. Similarly to the correction $\kappa$ of the muon $g-2$ in Eq.~\eqref{eq:gmin2_non-univ}, 
the electron EDM is induced as
\begin{align}
\frac{d_e}{e}
&=  \frac{m_e}{2}\, {\rm Im} \left[ A^L_{11} - A^R_{11} \right] \notag \\
&=  (1 + \delta^{\rm 2loop}) \frac{\alpha_Y}{8\pi} 
M_1\mu\tan\beta \notag \\
&~~~ \times
\bigg[
\frac{m_\mu}{1+\Delta_\mu} 
{\rm Im} \left[ (\delta_R)_{12} (\delta_L)_{12}^* \right]
\left(F_{1,1} - F_{1,2} - F_{2,1} + F_{2,2}\right)
\notag \\
&~~~~~~ + 
\frac{m_\tau}{1+\Delta_\tau} 
{\rm Im} \left[ (\delta_R)_{13} (\delta_L)_{13}^* \right]
\left(F_{1,1} - F_{1,3} - F_{3,1} + F_{3,3}\right) 
+ \cdots \bigg].
\label{eq:EDM}
\end{align}
Here, we assume $\arg(M_1\mu\tan\beta) =0$.
The omitted terms are suppressed by orders of $(\delta_L)_{ij}$, $(\delta_R)_{ij}$ or $m_\mu/m_\tau$. In the last term, $F_{1,1} \gg F_{1,3}, F_{3,1}, F_{3,3}$ is obtained when the staus are heavy. On the other hand, the right hand side vanishes when the slepton masses are universal, because the complex phases can be rotated away. Comparing Eq.~\eqref{eq:EDM} with Eq.~\eqref{eq:gmin2_non-univ}, we obtain
\begin{align}
\frac{d_e / e}{a_\mu({\rm SUSY})} \simeq 
\frac{1}{2 m_{\mu}}
{\rm{Im}}[ (\delta_R )_{13}  (\delta_L)^{ \ast}_{13} ]\,
\frac{m_{\tau}}{m_\mu}\frac{1 + \Delta_{\mu}}{1 + \Delta_{\tau}},
\label{eq:EDMg-2}
\end{align}
for the mass spectrum \eqref{eq:non-univ} with $R_L, R_R \gg 1$. It is independent of superparticle mass spectra except through $\Delta_{\mu}$ and $\Delta_{\tau}$. Thus, if the muon $g-2$ anomaly is solved by the SUSY contributions, EDM becomes sizable by the non-universal slepton mass. Similarly to LFV, this is valid even for large selectron and smuon masses. 

In the right panel of Fig.~\ref{fig:LFVCPV}, contours of the electron EDM are shown. Here, the SUSY contributions to the muon $g-2$ is fixed to be $a_\mu({\rm SUSY}) = 1 \times 10^{-9}$ with $M_{\rm soft} = 30\TeV$. The result is almost independent of superparticle mass spectra except for small corrections, $\Delta_\mu$ and $\Delta_\tau$. The following data are used,
\begin{itemize}
\item the current limit with the YbF molecule, $|d_e| < 1.05 \times 10^{-27}\,e{\rm cm}$ at 90\% CL~\cite{Hudson:2011zz} (solid line in the figure).
\item the sensitivity with the Fr atom, $|d_e| = 10^{-29}\,e{\rm cm}$~\cite{Sakemi:2011zz} (dashed line). The experiment with the ThO molecule could have a similar sensitivity by accumulating data, $|d_e| = 1 \times 10^{-28}/\sqrt{({\rm day})}\,e{\rm cm}$~\cite{Vutha:2009ux,Campbell:2013ipa}.
\item the sensitivity with the YbF molecule, $|d_e| = 10^{-30}\,e{\rm cm}$~\cite{Kara:2012ay} (dotted line). The experiment with the WN ion can probe down to, $|d_e| = 10^{-30}\,e{\rm cm}/{\rm day}$, where the systematic limit is at the level of $10^{-31}\,e{\rm cm}$~\cite{Kawall:2011zz}.
\end{itemize}
From the figure, it is found that EDM is sensitive to ${\rm{Im}}[ (\delta_R )_{13} (\delta_L)^*_{13} ]$ when the staus are heavier than the selectrons. The current experimental limit puts a constraint, ${\rm{Im}}[ (\delta_R )_{13} (\delta_L)^*_{13} ] < 6 \times 10^{-7}$ for $R \gg 1$ and $a_\mu({\rm SUSY}) = 1 \times 10^{-9}$. The sensitivity will be improved very well. The mixing will be able to be probed at the level of $10^{-10}$. Thus, if sleptons are neither discovered at LHC nor ILC in future, the model can be sensitively probed by EDM as well as LFV. If no signal will be observed, the CP violating phase must be suppressed very tightly in order to explain the muon $g-2$ anomaly by the SUSY contributions.


\section{Conclusion} \label{sec:conclusion}

SUSY is one of the most motivated candidates of the new physics. If the muon $g-2$ anomaly is solved by SUSY, some of the superparticles are relatively light. In this paper, we focused on the Bino--smuon contribution to the muon $g-2$. It is enhanced not only by $\tan\beta$ but also by large $\mu$. Consequently, it was shown that the smuon masses can be as large as $\sim 1\TeV$. We examined the phenomenology of the models, in which only the superparticles that are relevant for the muon $g-2$ are light. 

The analyses were categorized by the slepton mass spectrum. When the mass spectrum is universal among the flavors, it was found that the vacuum stability of the stau--Higgs potential restricts the smuon masses tightly. 
They are predicted to be within $330~(460)\GeV$ at the $1\sigma~(2\sigma)$ level of the muon $g-2$. 
It was shown that part of the parameter region is already excluded by LHC, and argued that such sleptons are expected to be studied at LHC or ILC in future. 

If the staus are (much) heavier than the smuons, the vacuum stability bound of the staus is relaxed. In this mass spectrum, the smuon masses are limited by the vacuum stability condition of the smuon--Higgs potential. It was found that they are less than $1.4~(1.9)\TeV$ at the $1\sigma~(2\sigma)$ level of the muon $g-2$. Such slepton masses exceed the LHC/ILC reach. Instead, the non-universal slepton mass spectrum generically predicts too large LFV and EDM. They originate in the non-diagonal SM Yukawa matrices in the model basis. Although the prediction depends on the flavor models, LFV and/or EDM is very likely to be sizable in wide models. If no sleptons are discovered at LHC nor ILC, the model is expected to be probed by LFV and EDM sensitively. 
 
So far, $\tan \beta = 40$ was chosen in the numerical analyses. As mentioned in Sec.~\ref{sec:setup}, the observables depend on $m_{\tilde\ell_{LR}}^2$, and $\tan \beta$ contributes to them only through it. Thus, the results are independent of a choice of $\tan \beta$ up to a correction to the vacuum stability condition (see Sec.~\ref{sec:vac}). Since the correction is small, the above conclusion does not change, even if $\tan \beta$ is varied, for instance, from 20 to 70.

In conclusion, we studied the SUSY models that explain the muon $g-2$ discrepancy with the Bino--smuon contributions. It was shown that they are expected to be probed by LHC/ILC and LFV/EDM complementarily in future.

\section*{Acknowledgements}
This work was supported by JSPS KAKENHI Grant No.~23740172 (M.E.), 22244021 (K.H.) and 25-10486 (T.Y.).
The work of T.K. is partially supported by  Global COE Program ``the Physical Sciences Frontier", MEXT, Japan. 
The work of T.Y. is supported in part by a JSPS Research Fellowship for Young Scientists, and
supported by an Advanced Leading Graduate Course for Photon Science grant.

\section*{Note Added}

While we are finalizing the paper, the article~\cite{Fargnoli:2013zda} was submitted to arXiv. 
The article studied non-decoupling two-loop contributions from heavy sfermions, 
which partially overlaps with the discussion on the correction to the muon $g-2$ via the Bino--muon--smuon coupling in Sec.~\ref{sec:g-2}.

\providecommand{\href}[2]{#2}
\begingroup\raggedright
\endgroup

\end{document}